\documentclass[twocolumn]{aastex63}
\usepackage[utf8]{inputenc}

\usepackage{amsmath}
\usepackage{hyperref}
\usepackage{color,soul}

\graphicspath{{./}{figures/}}

\begin{document}

\title{The transmission spectrum of WASP-17\,b from the optical to the near-infrared wavelengths: combining STIS, WFC3 and IRAC datasets}

\author{Arianna Saba}
\affiliation{Department of Physics and Astronomy, University College London, Gower Street, WC1E 6BT London, United Kingdom}
\author{Angelos Tsiaras}
\affiliation{INAF - Osservatorio Astrofisico di Arcetri, 
Largo E. Fermi 5, 50125 Firenze, Italy}
\affiliation{Department of Physics and Astronomy, University College London, Gower Street, WC1E 6BT London, United Kingdom}
\author{Mario Morvan}
\affiliation{Department of Physics and Astronomy, University College London, Gower Street, WC1E 6BT London, United Kingdom}
\author{Alexandra Thompson}
\affiliation{Department of Physics and Astronomy, University College London, Gower Street, WC1E 6BT London, United Kingdom}
\author{Quentin Changeat}
\affiliation{Department of Physics and Astronomy, University College London, Gower Street, WC1E 6BT London, United Kingdom}
\author{Billy Edwards}
\affiliation{AIM, CEA, CNRS, Universit\'e Paris-Saclay, Universit\'e de Paris, F-91191 Gif-sur-Yvette, France}
\affiliation{Department of Physics and Astronomy, University College London, Gower Street, WC1E 6BT London, United Kingdom}
\author{Andrew Jolly}
\affiliation{School of Physics, University of New South Wales, Sydney, NSW 2052, Australia}
\affiliation{Department of Physics and Astronomy, University College London, Gower Street, WC1E 6BT London, United Kingdom}
\author{Ingo Waldmann}
\affiliation{Department of Physics and Astronomy, University College London, Gower Street, WC1E 6BT London, United Kingdom}
\author{Giovanna Tinetti}
\affiliation{Department of Physics and Astronomy, University College London, Gower Street, WC1E 6BT London, United Kingdom}

\begin{abstract}
We present the transmission spectrum of the inflated hot-Jupiter WASP-17\,b, observed with the STIS and WFC3 instruments aboard the Hubble Space Telescope, allowing for a continuous wavelength coverage from $\sim$0.4 to $\sim$1.7 $\mu$m. Observations taken with IRAC channel 1 and 2 on the Spitzer Space Telescope are also included, adding photometric measurements at 3.6 and 4.5 $\mu$m. HST spectral data was analysed with Iraclis, a pipeline specialised in the reduction of STIS and WFC3 transit and eclipse observations. Spitzer photometric observations were reduced with the TLCD-LSTM method, utilising recurrent neural networks. The outcome of our reduction produces incompatible results between STIS visit 1 and visit 2, which leads us to consider two scenarios for G430L. Additionally, by modelling the WFC3 data alone, we can extract atmospheric information without having to deal with the contrasting STIS datasets. We run separate retrievals on the three spectral scenarios with the aid of TauREx\,3, a fully Bayesian retrieval framework. We find that, independently of the data considered, the exoplanet atmosphere displays strong water signatures and potentially, the presence of aluminium oxide (AlO) and titanium hydride (TiH). A retrieval that includes an extreme photospheric activity of the host star is the preferred model, but we recognise that such a scenario is unlikely for an F6-type star. Due to the incompleteness of all STIS spectral lightcurves, only further observations with this instrument would allow us to properly constrain the atmospheric limb of WASP-17\,b, before JWST or Ariel will come online.

\vspace{10mm} 
    
\end{abstract}

\section{Introduction}
% Present your topic and get the reader interested.
% Provide background or summarize existing research.
% Position your own approach.
% Detail your specific research problem.
% Give an overview of the paper's structure.
In recent years, the \textit{Hubble Space Telescope} has been at the forefront of exoplanet atmospheric characterisation. Thanks to its Wide Field Camera 3 (WFC3) and its Space Telescope Imaging Spectrograph (STIS), often combined with the InfraRed Array Camera (IRAC) aboard the \textit{Spitzer Space Telescope}, the atmosphere of hot-Jupiters \citep{sing2016continuum, tsiaras2018population, 10.1093/mnras/sty2544, skaf2020ares, pluriel2020ares}, ultra-hot-Jupiters \citep{kreidberg2018global, mikal2019emission, edwards2020ares, changeat2020kelt}, super-Earths \citep{tsiaras2016detection, tsiaras2019water, edwards2020hubble, mugnai2021ares}, sub-Neptunes \citep{guo2020updated, guilluy2020ares} and hot-Saturns \citep{anisman2020wasp, carone2021indications} have been studied in transmission, emission and across their phase-curves \citep{knutson20123,stevenson2017spitzer, kreidberg2018global, arcangeli2019climate, irwin20202, feng20202d,  changeat2021exploration}. Although small rocky planets are difficult to probe with current spectrographs due to the limited signal to noise ratio of their gaseous envelope, atmospheric studies conducted on extrasolar gas giants reported the detection of various molecules, in particular water vapour. \par
Among the exoplanet population of inflated hot-Jupiters, WASP-17\,b \citep{2010ApJ...709..159A} is one of the least dense planets discovered so far, with a radius equal to 1.932 $\pm$ 0.053 R$_J$ and a mass 0.477 times that of Jupiter \citep{2012MNRAS.426.1338S}. The transiting planet, confirmed to possess a retrograde orbit \citep{2010ApJ...722L.224B, hebrard2011retrograde, simpson2011spin}, forms an angle of 148.5$^{\circ}$ $^{+5.1}_{-4.2}$ with respect to the rotation axis of its parent star \citep{triaud2010spin}. The considerable orbital obliquity of WASP-17\,b cannot be explained by inward migration \citep{lin1996orbital, ida2004toward} but could be rather due to an initial misalignment between the star and the protoplanetary disk  \citep{bate2010chaotic}, planet-planet scattering or the Kozai mechanism \citep{nagasawa2008formation}. These processes also induce the very small eccentricities reported in orbitally misaligned planets \citep{bourrier2018orbital, johnson2009third, winn2009spin}: in fact, \cite{2012MNRAS.426.1338S} were not able to find any evidence of the orbital eccentricity of WASP-17\,b. The exoplanet takes 3.7 days to orbit its host star \citep{2012MNRAS.426.1338S}, a F6-type star with V$_{mag}$=11.6 \citep{2010ApJ...709..159A}. \par
Given the planet's very large atmospheric scale height \citep{2010ApJ...709..159A}, transmission spectroscopy studies have been extremely successful. From the ground, sodium was detected at optical wavelengths by \cite{2011MNRAS.412.2376W} and \cite{2012MNRAS.426.2483Z} using the VLT, and with the Magellan Telescope by \cite{2018A&A...618A..98K}. At low-resolution, the spectrograph FORS2 mounted on the VLT detected potassium in the atmosphere of WASP-17\,b with a 3$\sigma$ significance and water absorption \citep{2016A&A...596A..47S}. From space, data taken with HST in the optical and near-infrared and with Spitzer at  3.6 $\mu$m and 4.5 $\mu$m, report water absorption and alkali features, with little to no clouds/hazes \citep{sing2016continuum}. A more recent atmospheric study on WASP-17\,b \citep{alderson2022comprehensive} further confirms the presence of H$_2$O in addition to CO$_2$ absorption in the infrared. Lastly, the study conducted by \cite{barstow2016consistent} reports the presence of water at solar abundances in the limb of WASP-17\,b, possibly accompanied by CO signatures. However, contrary to the results by \cite{sing2016continuum}, their best atmospheric model includes scattering aerosols at relatively high altitudes, up to 1 mbar.

Here we present the spectroscopic analysis performed on transit observations taken with STIS (gratings G430L and G750L) and WFC3 (grisms G102 and G141) onboard the \textit{Hubble Space Telescope} and with IRAC channel 1 and 2 of the \textit{Spitzer Space Telescope}. The combination of the aforementioned instruments allows us to investigate the spectrum of WASP-17\,b from the optical to the near infrared wavelengths (0.4 - 5 $\mu$m).

\begin{table}[ht!]
    \centering
    \caption{Stellar and planetary parameters for WASP-17\,b used for the data reduction with Iraclis and TLCD-LSTM, and the atmospheric modelling with TauREx\,3.}
    \begin{tabular}{cc}\hline\hline
    \multicolumn{2}{c}{Stellar \& Planetary Parameters}\\ \hline\hline
    $T_{*}$ [K] & 6550$\pm$80 $^{*}$\\
    $R_{*}$ [R$_\odot$] & 1.572$\pm$0.056 $^{*}$\\
    $M_{*}$ [M$_\odot$] & 1.306$\pm$0.026 $^{*}$\\
    $\log_{10}(g)$ [cm/s$^{2}$] & 4.161$\pm$0.026 $^{*}$\\
    $[\text{Fe/H}]$ & -0.19$\pm$0.09 $^*$\\
    a/R$_*$ & 7.27$^{+0.21}_{-0.44}$ \\
    %a/R$_*$ & 7.03$\pm$0.15 $^\ddagger$\\
    $e$ & 0 (fixed)\\
    $i$ [$^{\circ}$] &  87.96$^{+1.34}_{-1.56}$ \\
    %$i$ [$^{\circ}$] &  87.1$\pm$0.6 $^\ddagger$\\
    $\omega$ &  0 (fixed) \\
    $P$ [days] & 3.7354845$\pm$1.9$\times$10$^{-6}$ $^\dagger$\\
    $T_0$ [BJD$_{TDB}$] & 2454592.8015$\pm$0.0005 $^\dagger$\\
    $R_{p}/R_{*}$ & 0.12229$^{+0.00017}_{-0.00013}$\\
    $M_{p}$ [M$_\text{J}$] & 0.477$\pm$0.033 $^\dagger$\\
    $R_{p}$ [R$_\text{J}$] & 1.932$\pm$0.053 $^\dagger$ \\ \hline
        \multicolumn{2}{c}{$^*$\cite{anderson2011thermal}}\\ 
        \multicolumn{2}{c}{$^\dagger$\cite{2012MNRAS.426.1338S}}\\
        %\multicolumn{2}{c}{$^\ddagger$\cite{refId0}}\\
        %\multicolumn{2}{c}{$^\neg$\cite{2017AJ....153..136S}}\\
        %\multicolumn{2}{c}{$^\dashv$\cite{2019AJ....158..138S}}\\
        %\multicolumn{2}{c}{$^\dashv$\cite{2017ApJ...834...50B}}\\
        \hline \hline
    \end{tabular}
    \label{tab:iraclis_params}
\end{table}

\section{Methodology}
\subsection{HST data processing}

\subsubsection{WFC3}

We downloaded the HST/WFC3 raw spectroscopic and spatially scanned observations of WASP-17\,b from the Mikulski Archive for Space Telescopes (MAST), as part of the HST Proposal 14918 (P.I. Hannah Wakeford).
In our study we decided to not include the HST/WFC3 dataset from proposal 12181, obtained in staring mode, because of its low S/N. The data would produce error bars 6 to 7 times bigger compared to the other WFC3 observations, obtained in spatial scanning mode.

Two transits of the exoplanet were observed with the grisms G102 and G141, which cover the wavelengths between 0.8 - 1.1 $\mu$m and 1.1 - 1.7 $\mu$m respectively. We employed Iraclis \citep{tsiaras2016new}, an open-source reduction pipeline for WFC3\footnote{\url{https://github.com/ucl-exoplanets/Iraclis}}, to reduce the data and extract the lightcurves by performing the following steps: zero-read subtraction, reference pixels correction, non-linearity correction, dark current subtraction, gain conversion, sky background subtraction, calibration, flat-field correction, bad pixels and cosmic rays correction, and lightcurve extraction. 

The WFC3 detector induces two different time-dependent systematics to the lightcurve: one exponential, at the beginning of each HST orbit, and one linear throughout the visit. Iraclis corrects for these systematics by fitting a transit fit model $F(t)$, computed using Pylightcurve, on the white lightcurve. $F(t)$ is a function of the limb-darkening coefficients, $R_p/R_*$ and the orbital parameters $T_0, P, i, a/R_*, e, \omega$. The transit fit function is then multiplied by a normalisation factor $n_w$ and a function $R(t)$ containing a linear and an exponential term as follows:
\begin{equation}
    R(t) = (1 - r_a(t-T_0))(1-r_{b1} e^{-r_{b2}(t-t_0)}) \ . 
\label{eq:white_lightcurve}
\end{equation}

\begin{figure}[htp]
    \centering
    \includegraphics[width=0.4\textwidth, height=0.25\textheight]{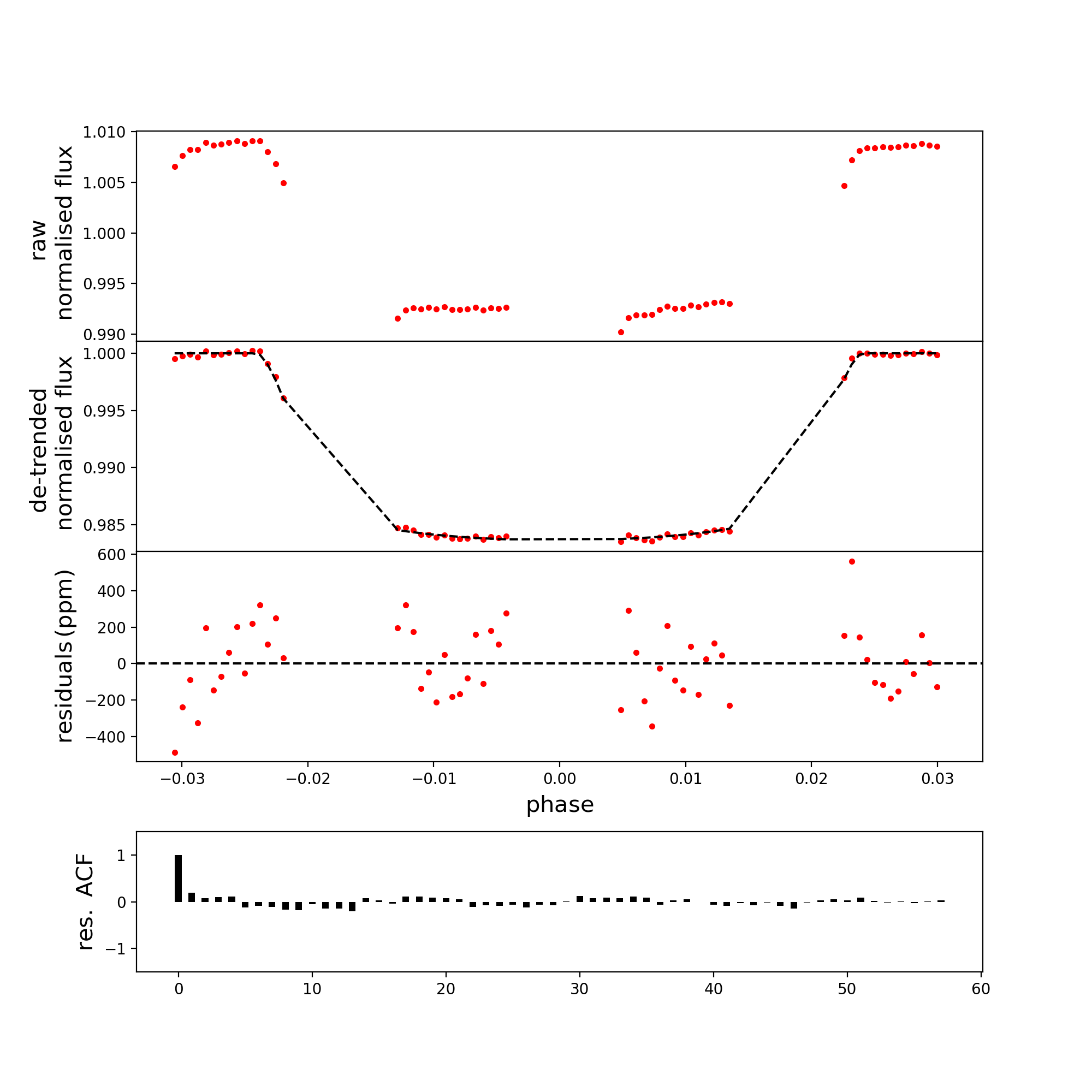}
    \includegraphics[width=0.47\textwidth, height=0.5\textheight]{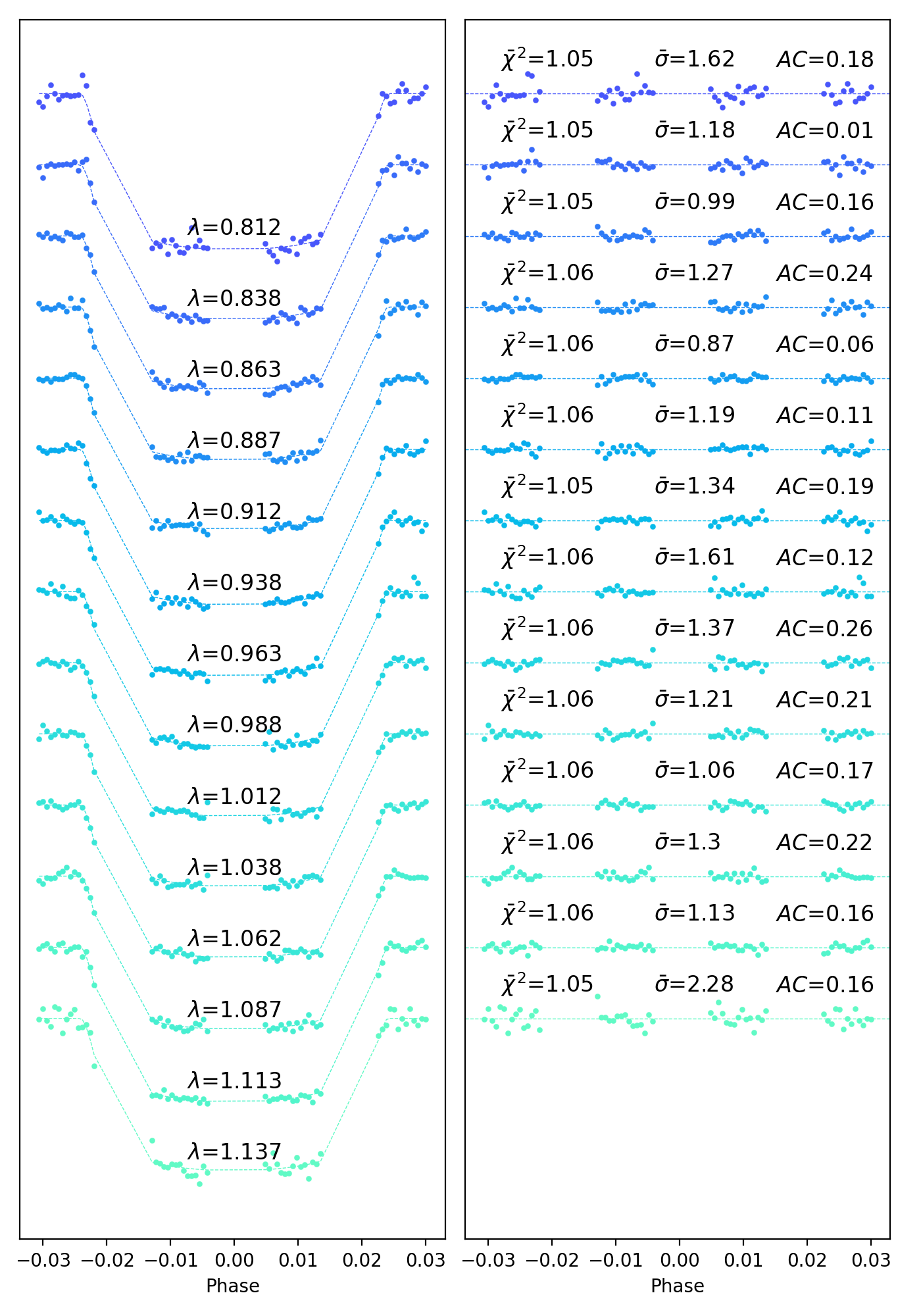}
    \caption{Top: white lightcurve for the HST/WFC3 G102 transmission observations of WASP-17\,b. First panel: raw lightcurve, after normalisation. Second panel: lightcurve, divided by the best fit model for the systematics. Third panel: residuals for best-fit model. Fourth panel: auto-correlation function of the residuals. Bottom: spectral lightcurves fitted with Iraclis for the HST/WFC3 G102 transmission spectra where, for clarity, an offset has been applied. Left panel: the detrended spectral lightcurves with the best-fit model plotted. Right panel: fitting residuals with values for the Chi-squared ($\chi^2$), the standard deviation of the residuals with respect to the photon noise ($\bar \sigma$) and the auto-correlation (AC) of the residuals.}
    \label{fig:transit_curve_g102}
\end{figure}

\begin{figure}[htp]
    \centering
    \includegraphics[width=0.4\textwidth, height=0.25\textheight]{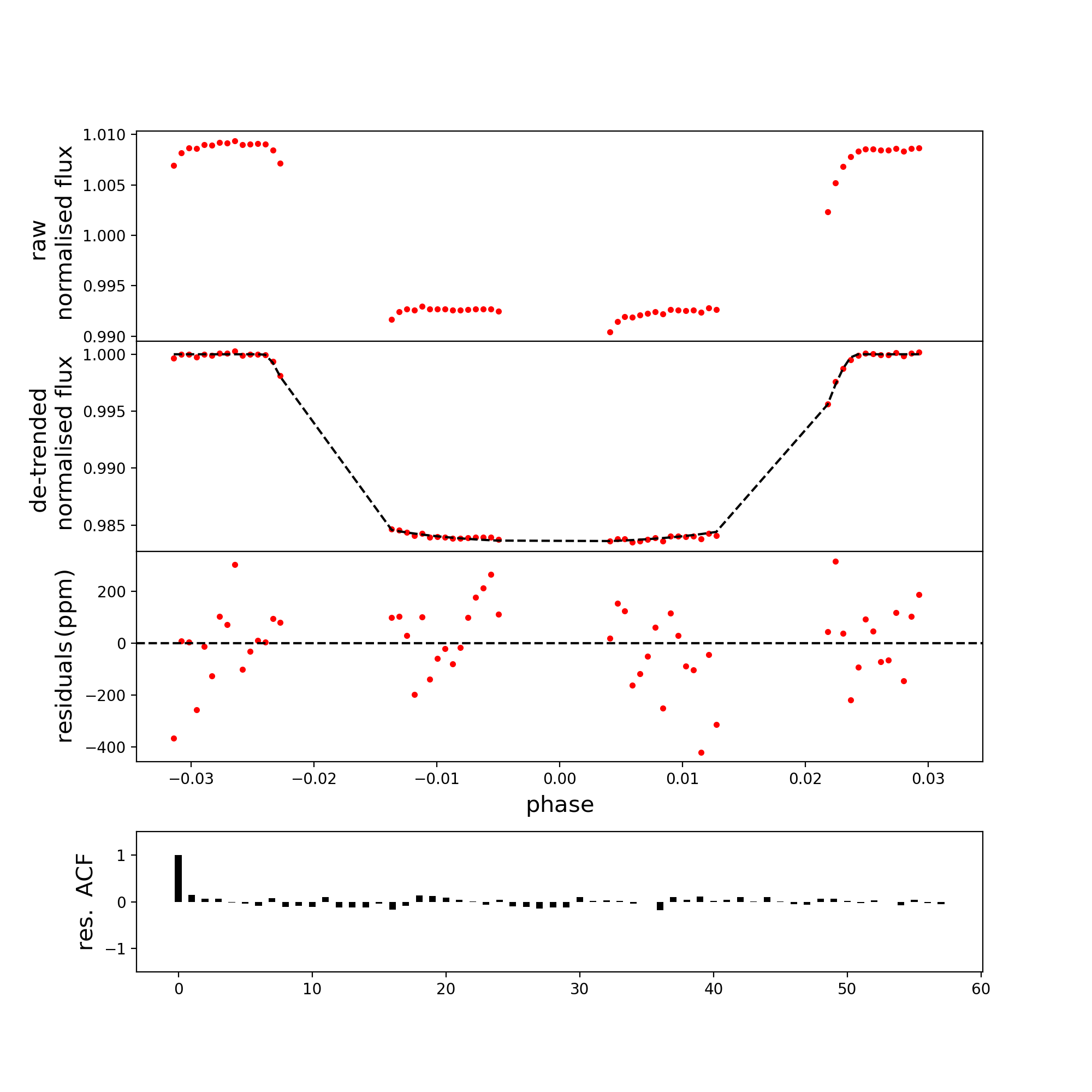}
    \includegraphics[width=0.47\textwidth, height=0.5\textheight]{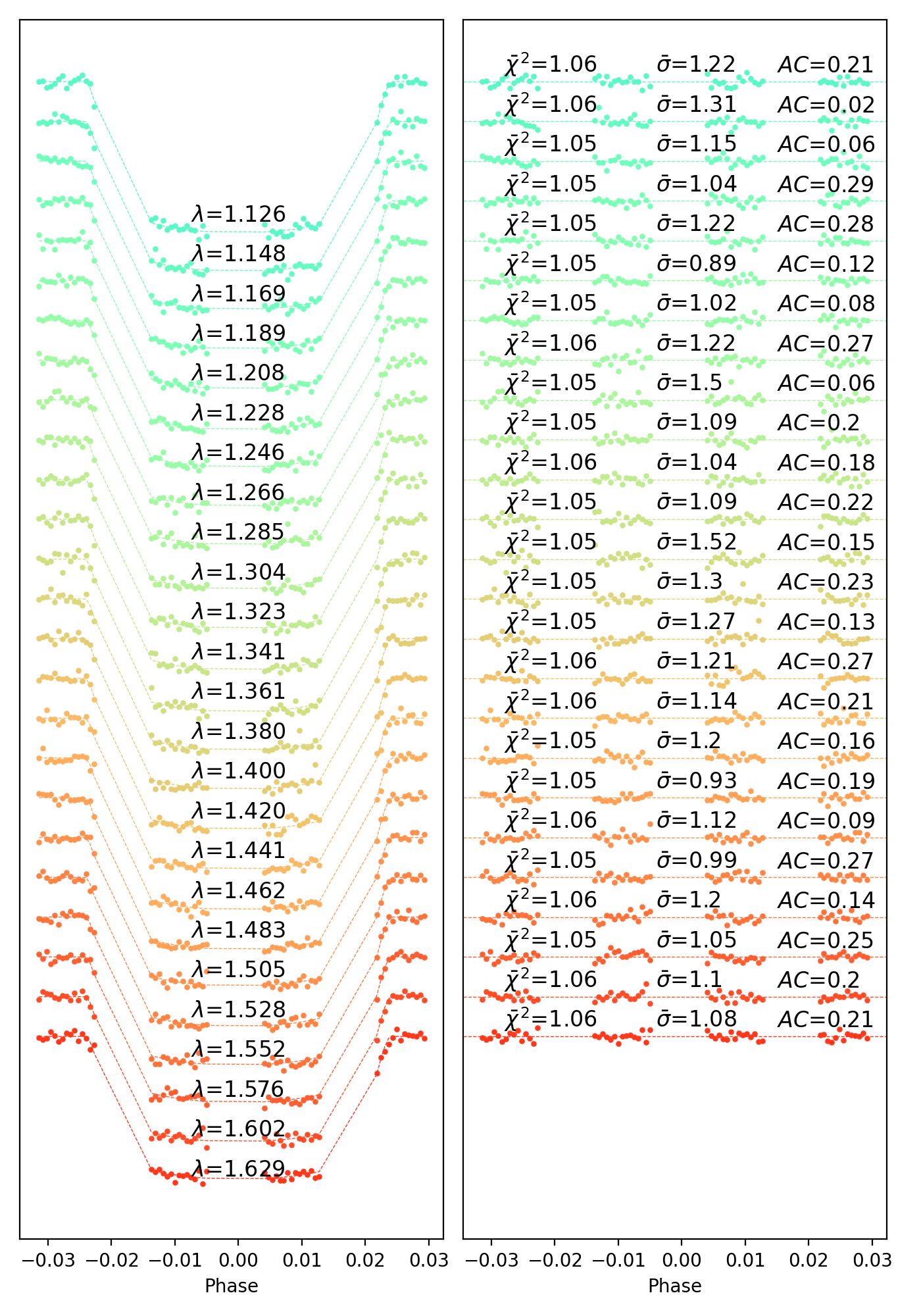}
    \caption{Top: white lightcurve for the HST/WFC3 G141 transmission observations of WASP-17\,b. First panel: raw lightcurve, after normalisation. Second panel: lightcurve, divided by the best fit model for the systematics. Third panel: residuals for best-fit model. Fourth panel: auto-correlation function of the residuals. Bottom: spectral lightcurves fitted with Iraclis for the HST/WFC3 G141 transmission spectra where, for clarity, an offset has been applied. Left panel: the detrended spectral lightcurves with the best-fit model plotted. Right panel: fitting residuals with values for the Chi-squared ($\chi^2$), the standard deviation of the residuals with respect to the photon noise ($\bar \sigma$) and the auto-correlation (AC) of the residuals.}
    \label{fig:transit_curve_g141}
\end{figure}

Here, $t$ is time, $T_0$ is the mid-transit time, $t_0$ is the time when each orbit starts, $r_a$ is the slope of the linear term and ($r_{b1}$, $r_{b2}$) are the coefficients of the exponential ramp.
The spectral lightcurves are firstly fitted using Equation\,\ref{eq:white_lightcurve}. Then, they are divided by the white lightcurve and further fitted via 
\begin{equation}
    n_{\lambda}(1+\chi_{\lambda}(t-T_0))(F(\lambda,t)/F_w(t)) \,
\end{equation}
where $n_{\lambda}$ is a wavelength-dependent normalisation factor, $\chi_{\lambda}$ is the coefficient of the wavelength-dependent linear slope, $t$ is time, $T_0$ is the mid-transit time, $F(\lambda,t)$ is the wavelength-dependent transit model and $F_w(t)$ is the best-fit model on the white lightcurve.

Further information on Iraclis can be found in \cite{tsiaras2016new, tsiaras2016detection, tsiaras2018population}.

\subsubsection{STIS}

We downloaded the HST/STIS raw spectroscopic observations of WASP-17\,b from the Mikulski Archive for Space Telescopes (MAST), as part of the HST Proposal 12473 (P.I. David Sing).

Three transits of the exoplanet were observed with the gratings G430L (two transits) and G750L (one transit), which cover the wavelengths between 0.3 - 0.57 $\mu$m and 0.5 - 0.94 $\mu$m respectively. For the analysis of these datasets we used again Iraclis but adapted most of the reduction steps to STIS. These steps are: bias level subtraction, bias correction, image subtraction, dark image subtraction, flat field correction, calibration, background subtraction, bad pixels and cosmic rays correction, and lightcurve extraction. The first five steps were performed following the recipes described in the STIS Handbook \citep{bostroem2011stis}, while the rest were implemented in the same way as for the WFC3 staring-mode observations. For the extraction we used 6-pixel-wide apertures along the cross-dispersion direction and smoothed aperture edges along the dispersion direction (the smoothing factors were 5 \AA  for the G430L grating and 10 \AA for the G750L grating, corresponding to approximately two pixels in each case).

\subsection{HST lightcurve modelling}

Table\,\ref{tab:iraclis_params} summarises the parameters used for the lightcurve modelling. For these parameters we consulted the Exoplanet Characterisation Catalogue developed as part of the ExoClock project \citep{Kokori_2021}. In all the lightcurve modelling steps we assumed a circular orbit and a fixed period for the planet, while the limb-darkening effect was modelled using the Claret 4-coefficient law \citep{Claret2000} and the ExoTETHyS \citep{morello2020exotethys} package which takes into account the stellar parameters (Table\,\ref{tab:iraclis_params}) and the response curve of the instrument.

\begin{figure}[htp]
    \centering
    \includegraphics[width=0.47\textwidth, height=0.3\textheight]{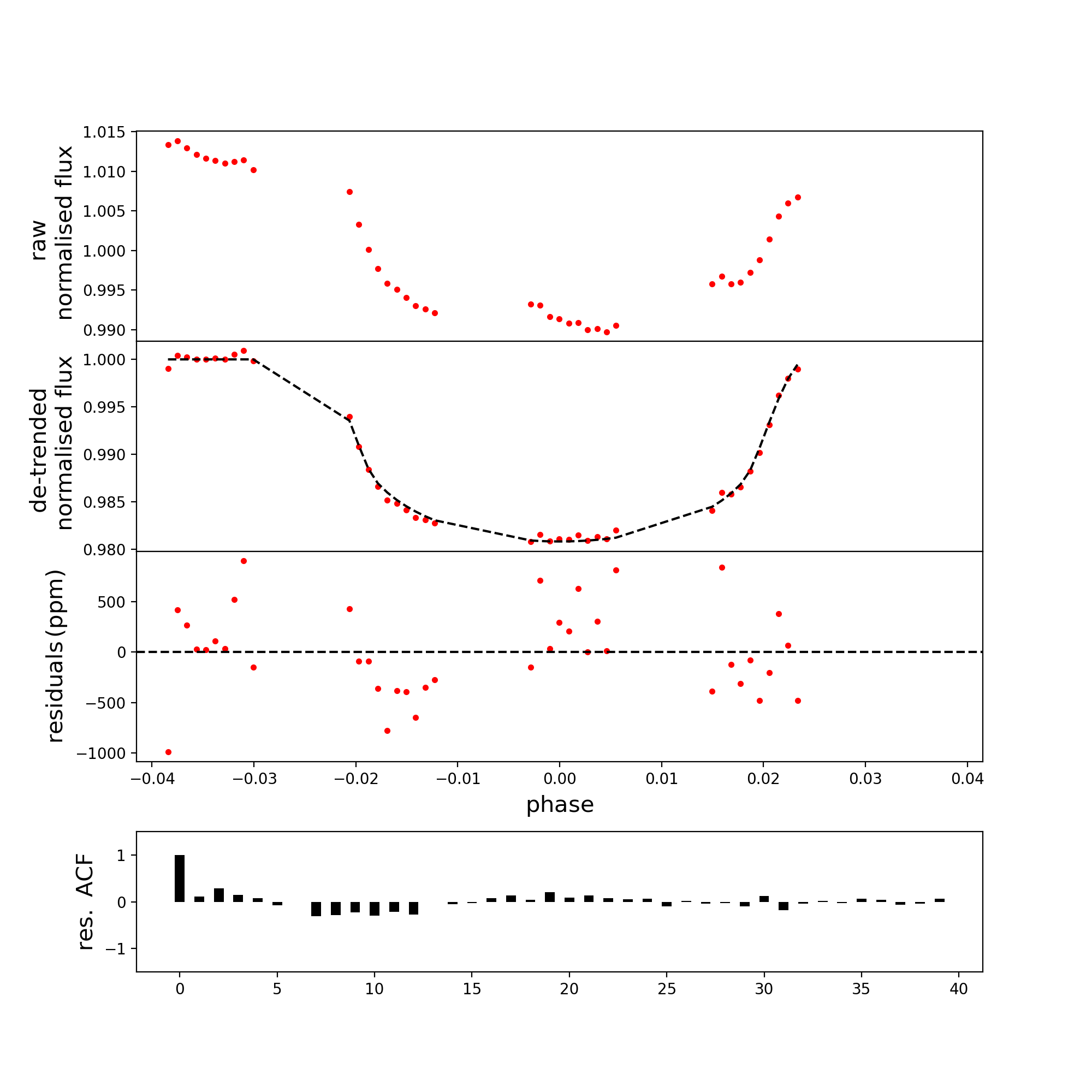}
    \includegraphics[width=0.47\textwidth, height=0.45\textheight]{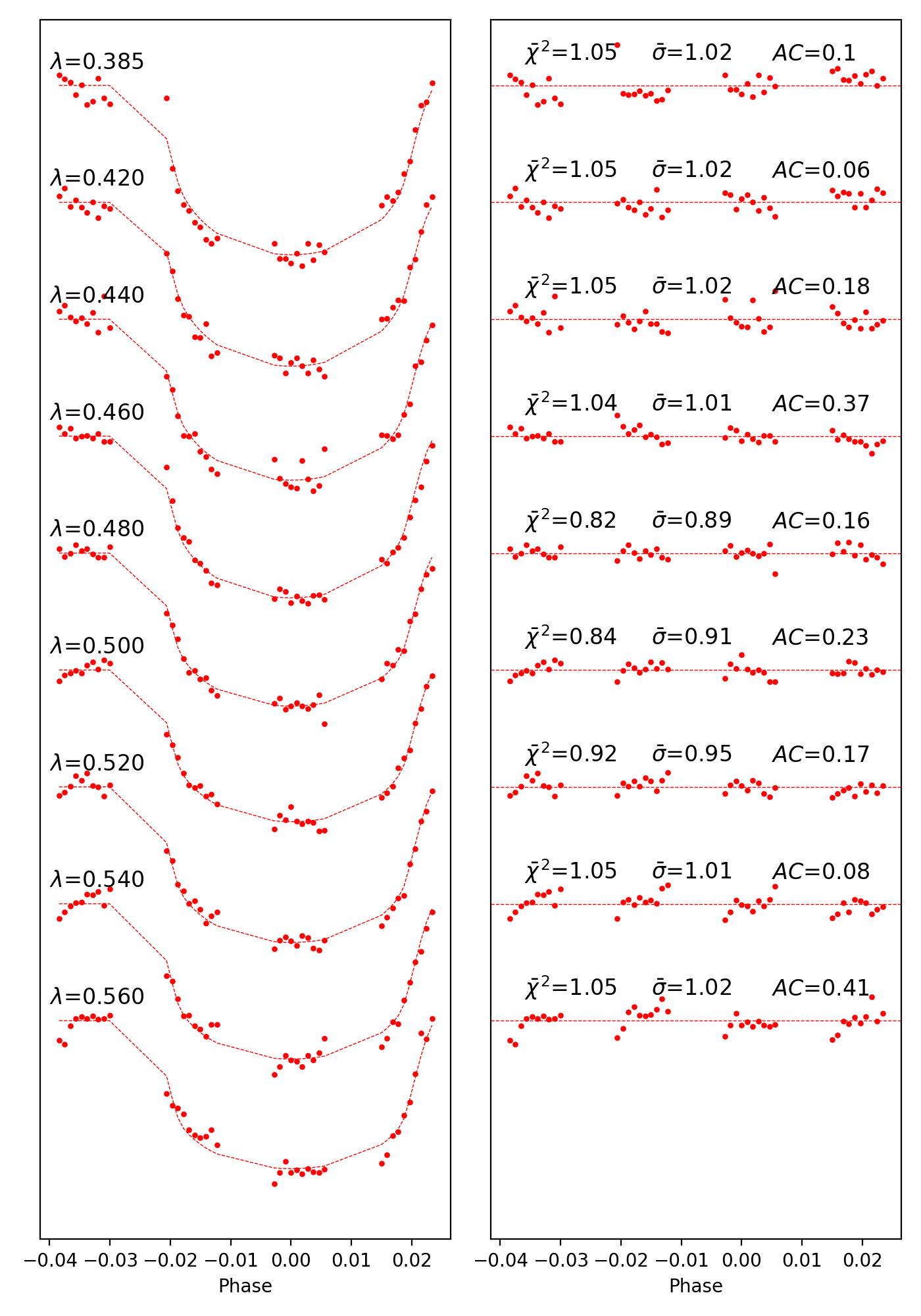}
    \caption{Top: white lightcurve for the HST/STIS G430L transmission observation taken on 2012-06-08 of WASP-17\,b. First panel: raw lightcurve, after normalisation. Second panel: lightcurve, divided by the best fit model for the systematics. Third panel: residuals for best-fit model. Fourth panel: auto-correlation function of the residuals. Bottom: spectral lightcurves fitted with Iraclis for the same observation where, for clarity, an offset has been applied. Left panel: the detrended spectral lightcurves with the best-fit model plotted. Right panel: fitting residuals with values for the Chi-squared ($\chi^2$), the standard deviation of the residuals with respect to the photon noise ($\bar \sigma$) and the auto-correlation (AC) of the residuals.}
    \label{fig:transit_curve_stis1}
\end{figure}

\begin{figure}[htp]
    \centering
    \includegraphics[width=0.47\textwidth, height=0.3\textheight]{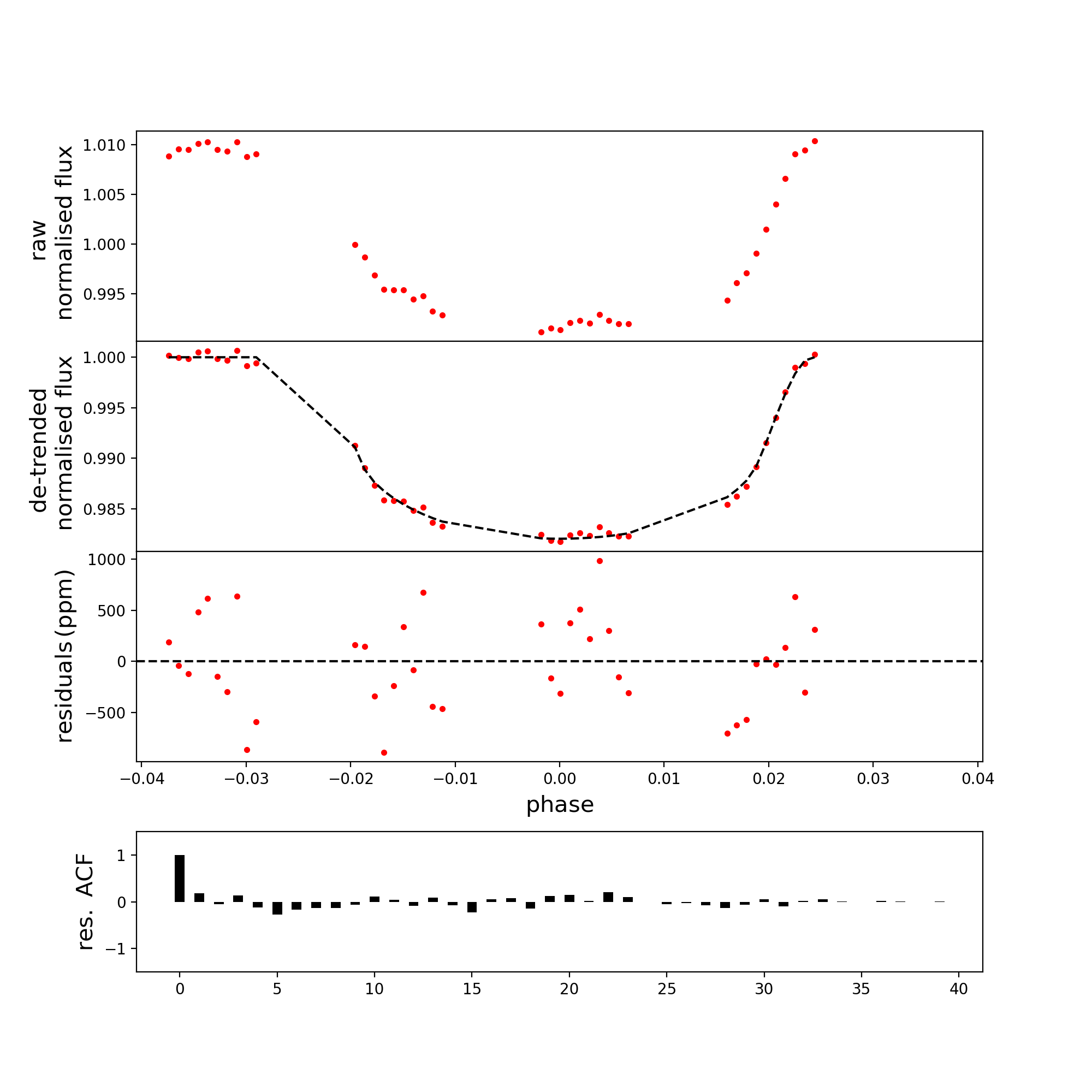}
    \includegraphics[width=0.47\textwidth, height=0.45\textheight]{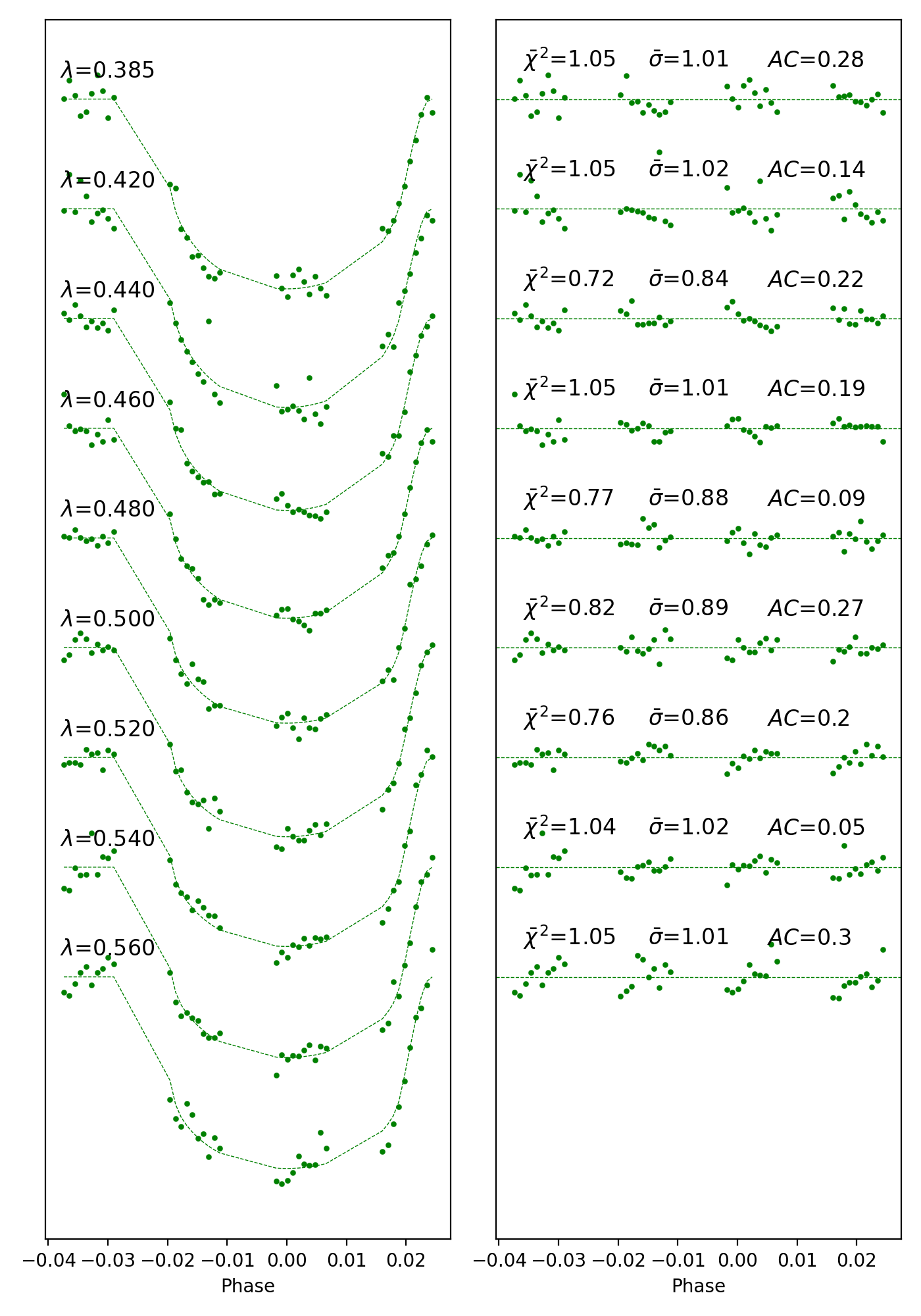}
    \caption{Top: white lightcurve for the HST/STIS G430L transmission observation taken on 2013-03-15 of WASP-17\,b. First panel: raw lightcurve, after normalisation. Second panel: lightcurve, divided by the best fit model for the systematics. Third panel: residuals for best-fit model. Fourth panel: auto-correlation function of the residuals. Bottom: spectral lightcurves fitted with Iraclis for the same observation where, for clarity, an offset has been applied. Left panel: the detrended spectral lightcurves with the best-fit model plotted. Right panel: fitting residuals with values for the Chi-squared ($\chi^2$), the standard deviation of the residuals with respect to the photon noise ($\bar \sigma$) and the auto-correlation (AC) of the residuals.}
    \label{fig:transit_curve_stis2}
\end{figure}

\begin{figure}[htp]
    \centering
    \includegraphics[width=0.47\textwidth, height=0.3\textheight]{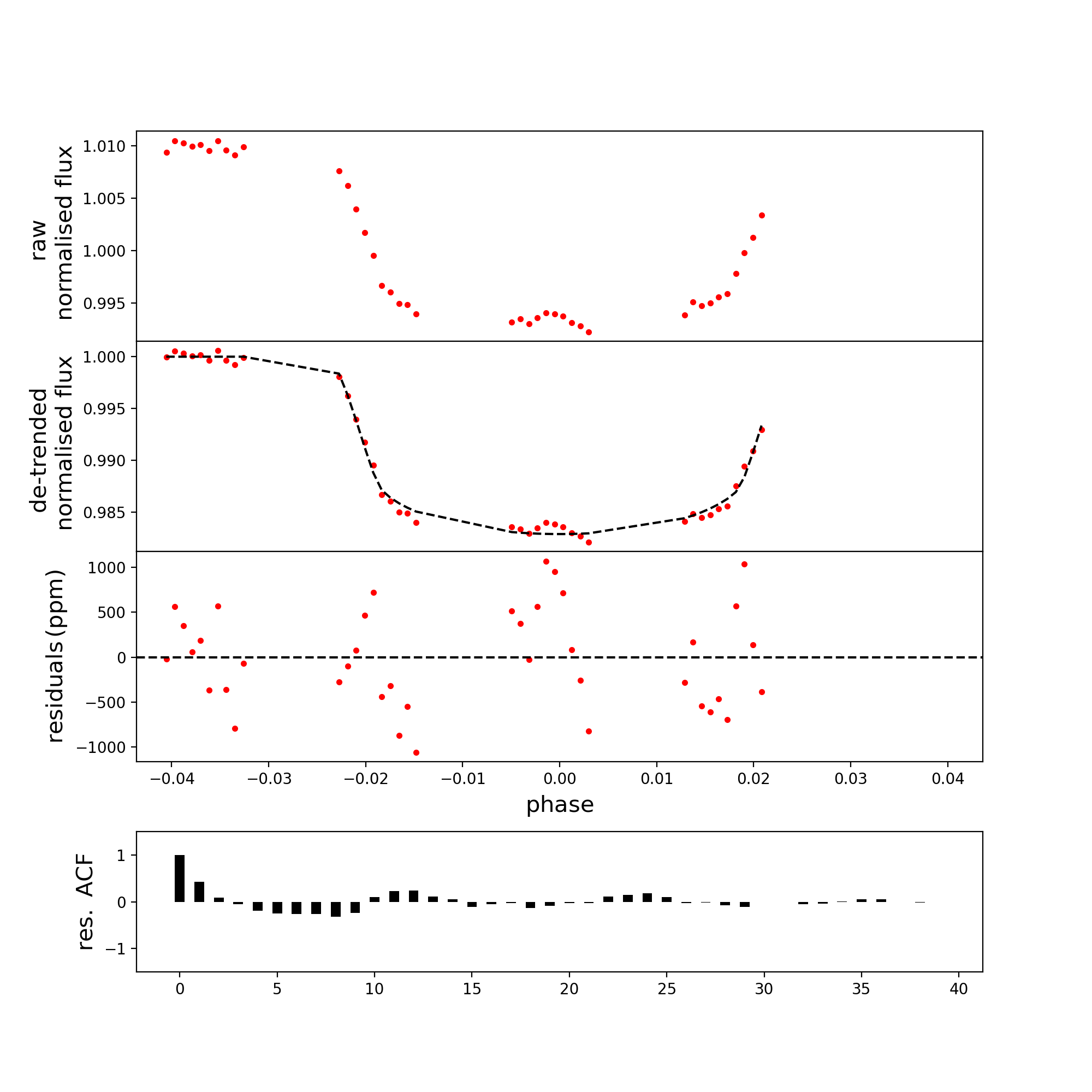}
    \includegraphics[width=0.47\textwidth, height=0.45\textheight]{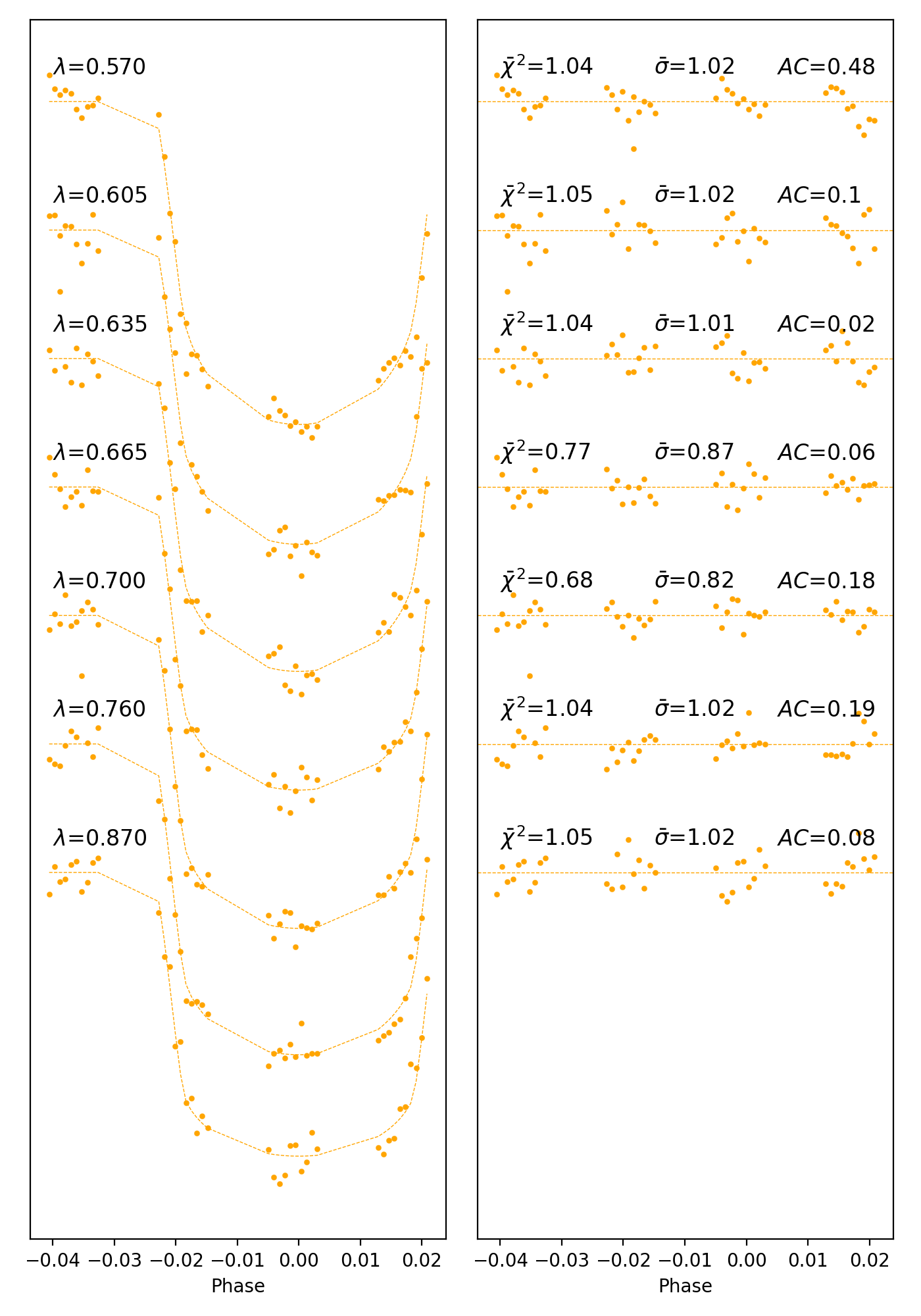}
    \caption{Top: white lightcurve for the HST/STIS G750L transmission observation taken on 2013-03-19 of WASP-17\,b. First panel: raw lightcurve, after normalisation. Second panel: lightcurve, divided by the best fit model for the systematics. Third panel: residuals for best-fit model. Fourth panel: auto-correlation function of the residuals. Bottom: spectral lightcurves fitted with Iraclis for the same observation where, for clarity, an offset has been applied. Left panel: the detrended spectral lightcurves with the best-fit model plotted. Right panel: fitting residuals with values for the Chi-squared ($\chi^2$), the standard deviation of the residuals with respect to the photon noise ($\bar \sigma$) and the auto-correlation (AC) of the residuals.}
    \label{fig:transit_curve_stis3}
\end{figure}

At a first stage, the initial orbit of each HST visit was discarded since it presents more significant wavelength-dependent systematics compared to the subsequent orbits. We then analysed only the white lightcurves (broad-band) and let the semi-major axis, the inclination, the mid-transit time and the $R_p/R_*$ as free parameters, together with a number of parameters for the systematics implemented in Iraclis.

To ensure consistency between the analysis of the different datasets, we estimated again the inclination and the semi-major axis of the planet using all the available data. We combined the lightcurves extracted from the five HST observations, together with five TESS \citep{ricker2014transiting} observations. The updated parameters are reported in Table\,\ref{tab:iraclis_params}.

Finally, we analysed all the white and spectral HST lightcurves fixing the values for the semi-major axis and the inclination. Figures \ref{fig:transit_curve_g102}, \ref{fig:transit_curve_g141}, \ref{fig:transit_curve_stis1},
\ref{fig:transit_curve_stis2} and
\ref{fig:transit_curve_stis3} show the fitted white and spectral lightcurves for all the HST observations while Table\,\ref{tab:stis_data} and \ref{tab:wfc3_data} report the transmission data for STIS and WFC3 respectively.

\subsection{Spitzer data analysis}
Spitzer data is available on the IRSA online archive. We downloaded the WASP-17\,b transit observations taken with IRAC channel 1 and 2 in 2013 (P.I. Jean-Michel Desert, programme 90092). The two observations were analysed separately by employing the lightcurve detrending method named TLCD-LSTM\footnote{\url{https://github.com/ucl-exoplanets/deepARTransit}} developed by \cite{morvan2020detrending}. This approach uses a Long Short-Term Memory network \citep{hochreiter1997long} to predict the transit lightcurve without the need for any prior assumptions on the noise or transit shape. The only assumption regarding the systematics is that the associated correlated noise can be inferred from the out-of-transit lightcurve and the centroid time series. First, photometric lightcurves are extracted by using a circular aperture around the center of light. We tested a grid of radii apertures ranging from 2 to 4 pixels, by increment of 0.25 pixels. We then selected the aperture which maximised the signal-to-noise ratio, i.e. 3.25 pixels. We subtracted the background light computed at each time step using the median flux in each image after excluding a disk of radius 15 pixels around the source centre.

\begin{figure}[htp]
    \centering
    \includegraphics[width=0.47\textwidth]{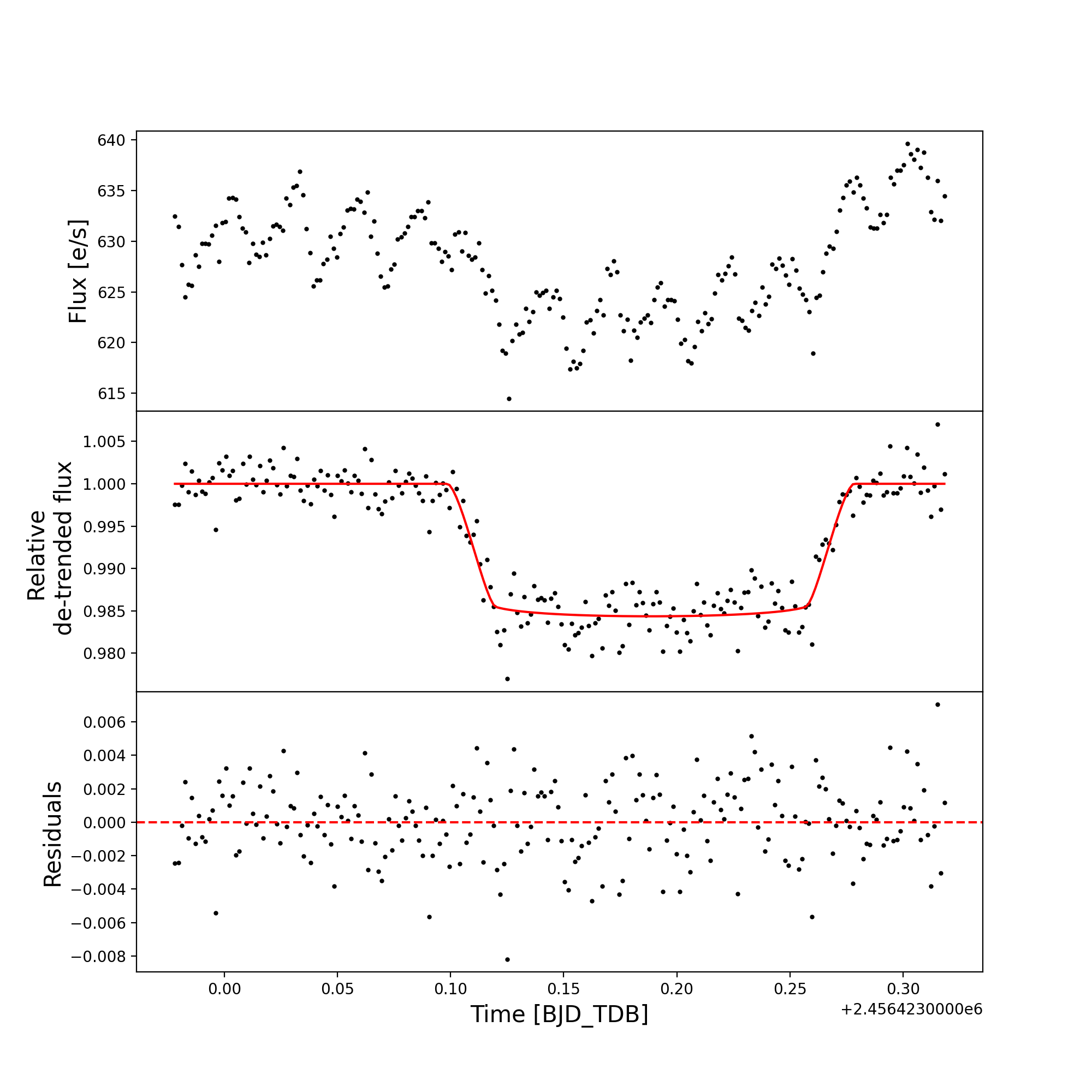}
    \includegraphics[width=0.47\textwidth]{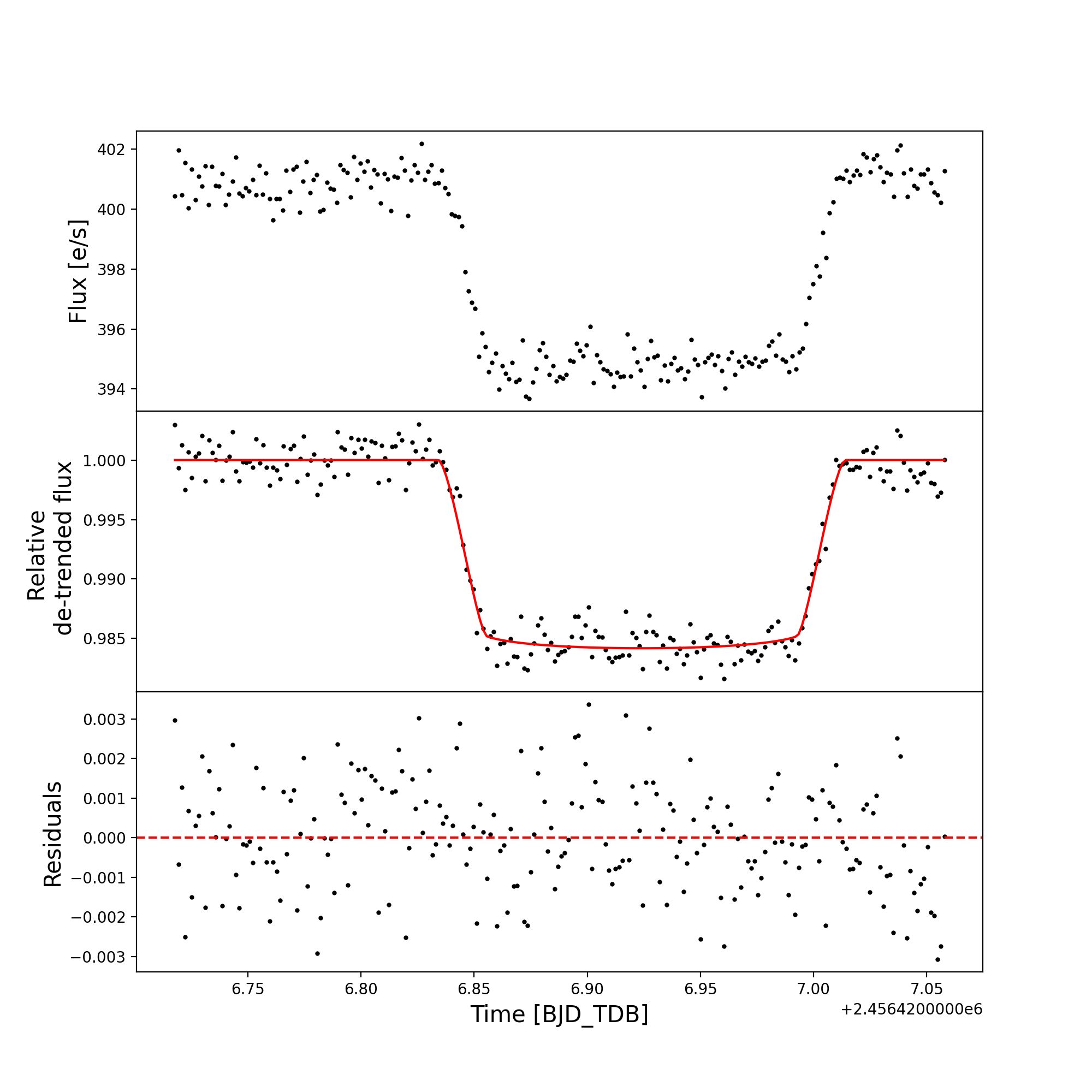}
    \caption{Raw lightcurves (first panel), detrended lightcurves (second panel) and residuals (third panel) for IRAC observations of WASP-17\,b with channel 1 (top) and 2 (bottom).}
    \label{fig:irac_detrended}
\end{figure}

The neural network was then trained to learn the temporal variability on the pre-ingress and post-egress parts of the photometric lightcurve.
While the stellar flux is masked during transit and then predicted autoregressively, the 2-D gaussian centroid time series is used as a covariate time series during the whole observation. This allows to leverage its correlations with the photometric lightcurve and infer the systematics during the transit while excluding the planetary signal.
However, intermediate transit fits are performed during training on the temporary detrended lightcurve in order to keep track of the model's progress and provide an additional stopping criterion. A first run is done with large margins around the expected ingress and egress times to ensure that the transits do not overlap with the training ranges. The margins are then refined after this first run to include 105\% of the transit duration centered on the mid-transit time. In practice, the results are very stable and unaffected by the chosen margins provided these remain low (under $\approx10\%$ of the transit duration). Several architecture and learning parameters are then tested. In both cases we find that 2 LSTM layers of 512 units and 10\% dropout trained with an Adam optimiser and decay rate of $\beta=0.95$ provide an optimal residual noise after the transit fit while avoiding overfitting. After the model was trained for 50 epochs, the best network - i.e. with the lowest residual noise - is saved along with the corresponding detrended lightcurve, obtained by dividing the raw lightcurve by the network's prediction. A final transit fit is performed using a Markov Chain Monte Carlo procedure embedded in the PyLightcurve package \citep{tsiaras2016new}. Both during and after training, the fixed planetary parameters used are those listed in Table\,\ref{tab:iraclis_params} or computed using the ExoTETHyS open package \citep{morello2020exotethys} while the mid-transit time and $R_p/R_*$ were let free. The full list of parameters used to train the network is available in Table\,\ref{tab:spitzer_network_params}.

The raw and detrended Spitzer lightcurves are shown in Figure \ref{fig:irac_detrended} (top: IRAC channel 1; bottom: IRAC channel 2) and the corresponding transmission data is available in Table \ref{tab:spitzer_data}.

\begin{figure*}[htp]
    \centering
    \includegraphics[width=\textwidth]{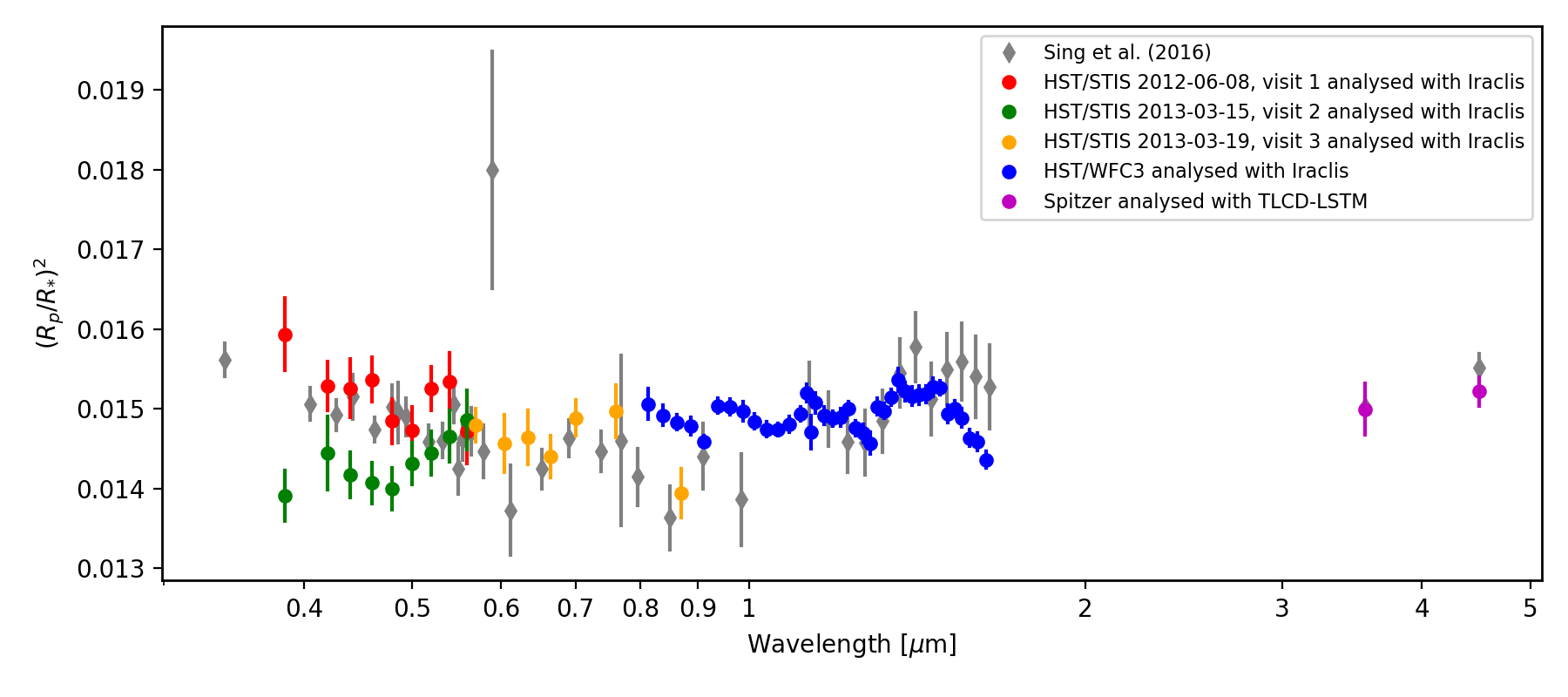}
    \caption{Transmission spectrum of WASP-17\,b, constituted by HST/STIS, HST/WFC3 and Spitzer/IRAC channel 1 and 2 data analysed in this study compared to the data from \cite{sing2016continuum}.}
    \label{fig:complete_spectrum}
\end{figure*}

\subsection{Atmospheric Modelling}
The spectrum resulting from our data reduction process is shown in Figure\,\ref{fig:complete_spectrum}. Each STIS dataset is indicated with a different colour depending on the date the observation was taken on. WFC3 and Spitzer data are shown in blue and purple respectively, independently of the grism or channel employed. From the aforementioned plot we can easily notice divergent results coming from the two STIS/G430L observations, namely those taken on 2012-06-08 and 2013-03-15. For this reason we decided to model the observed spectrum in three separate scenarios. Case 1 includes STIS visits 2012-06-08 (visit 1) and 2013-03-19 (visit 3), WFC3 and Spitzer data while in Case 2 we aimed to model STIS visits 2013-03-15 (visit 2) and 2013-03-19 (visit 3), WFC3 and Spitzer datasets. Additionally, we investigated the spectrum constituted by the WFC3 observations only (Case 3), thus discarding the inconsistencies of the STIS observations.

In Figure\,\ref{fig:complete_spectrum} we compare our data with the analysis from \cite{sing2016continuum}. We observe that our STIS visit 1, STIS visit 3, WFC3/G141 and Spitzer data agree with the previous analysis by \cite{sing2016continuum}, finding however an inconsistency with the data point at 0.6 $\mu$m. Additionally, we notice that our STIS visit 2 data do not have a counterpart in the analysis by \cite{sing2016continuum}, although their study claims to have analysed all STIS observations available. We believe that their approach consisted in performing a weighted average of STIS visit 1+2, but we cannot be sure as this information is not included in their methodology.
 
In each spectral case, we modelled the atmosphere at the terminator of WASP-17\,b using the open-source atmospheric retrieval framework TauREx\,3 \citep{Al_Refaie_2021}. As it is customary for hot-Jupiters, we assumed this exoplanetary atmosphere to be mainly constituted by H$_2$ and He in a ratio of 0.17. %Using an isothermal profile in this study was supported by the data being of low quality. 
Given the short orbital period and an unconstrained eccentricity, we expect a tidally-locked exoplanet with an equilibrium temperature that remains constant throughout the whole orbit around the host star. Using the equation
\begin{equation}
    T_{p} = T_* \sqrt{\frac{R_*}{2a}} \left( \frac{1-A}{\epsilon} \right)^{\frac{1}{4}} \ ,
\label{eq:planet_temp}
\end{equation}
we estimated the equilibrium temperature to be $T_{p}$ = 1769 K when assuming $A$ = 0.1 and $\epsilon$ = 0.8. However, since we cannot ascertain the exact albedo and emissivity of the planet, we considered some extreme cases which produce a range where we expect the planetary temperature to fall. For instance, if we assume WASP-17\,b to possess an albedo of 0.5 and an emissivity of 0.7 (analogue to Jupiter), the planetary equilibrium temperature suddenly drops to $T_{p}$ = 1579 K. If instead we consider the case of a planet that reflects a tiny amount of incoming light ($A$ = 0.1) and re-emits some energy (possibly energy left over from the planet's formation process) with $\epsilon$ = 0.7, then the equilibrium temperature increases to $T_{p}$ = 1829 K. \par
As a result of the broad wavelength coverage of the combined observations and the high planetary equilibrium temperature \citep{sharp2007atomic}, we explored the presence of a variety of molecular opacities, ranging from the optical (TiO \citep{mckemmish2019exomol}, VO \citep{mckemmish2016exomol}, FeH \citep{dulick2003line, wende2010crires}, SiH \citep{yurchenko2018exomol}, TiH \citep{gharib2021exoplines}, AlO \citep{patrascu2015exomol}, K \citep{allard2019new} and Na \citep{allard2019new}), to the near-infrared and infrared absorbers (H$_{2}$O \citep{10.1093/mnras/sty1877}, CH$_{4}$ \citep{Yurchenko_2014}, CO \citep{Li_2015}, CO$_2$ \citep{2010JQSRT.111.2139R} and NH$_3$ \citep{coles2019exomol}), by employing the linelists from the ExoMol \citep{2016JMoSp.327...73T, chubb2021exomolop}, HITEMP \citep{rothman2010hitemp, rothman2014status}, HITRAN \citep{2016DPS....4842113G, rothman1987hitran} and NIST \citep{kramida2013critical} databases. %However, during our preliminary retrievals we could not constrain the presence of CH$_{4}$, CO, CO$_2$, NH$_3$, K, Na. %The reason being the carbon-bearing molecules possessing strong absorption features in the infrared range that cannot be probed with current instruments and the alkali features expected to show strong lines at particular wavelengths, that we were not able to retrieve. 

In our final retrievals we considered only the presence of H$_2$O, AlO, TiH, SiH, CO, CH$_4$ as trace gases. The molecular profile of each species was set to be constant at each atmospheric layer. Additionally, collision Induced Absorption (CIA) from H$_2$-H$_2$ \citep{PMID:21207941, Fletcher_2018} and H$_2$-He \citep{PMID:22299883}, Rayleigh scattering for all molecules and the presence of gray clouds and hazes modelled with the \cite{lee2013atmospheric} parameterisation were included. We also modelled the contamination of the host star on the spectrum of WASP-17\,b using the formalisation by \cite{rackham2018transit, rackham2019transit}. \par
%Given the steep downward and upward slope of STIS visit 1 and 2 observations, we additionally tested the presence of spots and faculae on the stellar photosphere, albeit a F6 type star like WASP-17 is not expected, by the current stellar models \citep{rackham2019transit}, to show an extreme activity. We modelled the spot temperature within the 4000 and 5000 K range, while the faculae temperature could adjust between 6600 and 7000 K. Both the spots and faculae coverage fraction bounds where set between 0.0 and 0.9 of the total stellar surface. 
We assumed a planetary atmosphere constituted by 100 layers in a plane-parallel geometry, uniformly distributed in log space between $10^{-5}$ and $10^{6}$ Pa. The temperature structure was modelled with an isothermal $T-p$ profile. The trace gases considered were allowed to vary freely between $10^{-12}$ and $10^{-1}$ in volume mixing ratio, while the planetary equilibrium temperature, $T_p$, could vary between 500 and 3000 K. Regarding the hazes, their top pressure $P_{Mie}^{top}$ ranged from $10^{-4}$ to $10^{6}$ Pa, the radius of the particles $R_{Mie}^{Lee}$ between $10^{-3}$ and 1 $\mu$m and the Mie cloud mixing ratio $\chi_{Mie}^{Lee}$ could change between -30 and -4 in log-space. Furthermore, we included a layer of gray clouds, whose top pressure could vary between $10^{-1}$ and $10^{6}$ Pa \citep{robinson2014common, kawashima2018theoretical, charnay2021formation}. Lastly, the planetary radius bounds where set between 1 and 2 R$_J$. The prior bounds employed for each fitted parameter are reported in Table\,\ref{tab:retrieval}.

We investigated the parameter space with the nested sampling algorithm MultiNest \citep{10.1111/j.1365-2966.2009.14548.x} with 750 live points and an evidence tolerance of 0.5.

\subsection{Stellar Modelling}
We attempted to explain the steep downward slope in the blue wavelengths of STIS visit 2 observation (see Figure\,\ref{fig:complete_spectrum}) by modelling the effect of a combination of unocculted stellar spots and faculae on the Case 2 transmission spectrum of WASP-17\,b. We accounted for solar size spots and faculae to be homogeneously distributed on the stellar surface. Current stellar models and observational evidence \citep{rackham2019transit, rackham2017access, mamajek2008improved, ciardi2011characterizing} show that the variability of F-dwarfs is $\sim$0.1\%, with a minimum of 0.07\% and a maximum of 0.36\%. This is one of the lowest among all stellar spectral types, with the stellar variability being inversely proportional to the temperature of the star \citep{mcquillan2014rotation}. Though generally FGK stars are mostly inactive and do not contaminate the transmission spectrum of the planets orbiting them, their active counter-parts do, and great care must be taken when analysing the planetary spectra that they pollute. \par
According to \cite{rackham2019transit}, for FGK-type stars the stellar spot temperature can be calculated as:
\begin{equation}
    T_{spot} \ [K] = 0.418 \ T_{phot} +1620 
    \label{eq:spots}
\end{equation}
and the faculae temperature as:
\begin{equation}
    T_{fac} \ [K] = T_{phot} + 100 \ ,
    \label{eq:faculae}
\end{equation}
with $T_{phot}$ being the temperature of the stellar photosphere, in our study equal to 6550$\pm$80 K. The spectral emission density of the heterogeneous star is modelled using the BT-Settl models \citep{allard2013progress} and the stellar spectral model grid is generated using the PHOENIX library \citep{husser2013new}. In essence, the spots and faculae are modelled as separate cooler and hotter stars respectively (Thompson et al.(in prep)). The stellar emission densities (SEDs) of the three surface components (spots, faculae and quiescent photosphere) are then combined in the ratio of their covering fractions to produce the observed disk-integrated stellar spectrum 
\begin{equation}
\begin{split}
    S_{star, \lambda} & = ((1-F_{spot}-F_{fac}) \times S_{phot, \lambda}) + \\
    & (F_{spot} \times S_{spot,\lambda}) + (F_{fac}\times S_{fac,\lambda}) \ ,
\end{split}
\end{equation}
where $S_{star,\lambda}$ is the flux of the heterogeneous star at a given wavelength, $S_{phot,\lambda}$, $S_{spot,\lambda}$ and $S_{fac,\lambda}$ are the spectra of the quiet photosphere, the starspots and the faculae at the same wavelength respectively and $F_{spot}$ and $F_{fac}$ are the covering fractions of the spots and faculae with respect to the observed stellar disk. \par
Compared to the nominal transit depth, the effect of an heterogeneous star produces a wavelength-dependent offset on the transmission spectrum given by
\begin{equation}
%\resizebox{0.4\textwidth}{!} %, height=0.02\textheight}
    \epsilon_{\lambda, s+f} = \frac{1}{1-F_{spot}\left(1-\frac{S_{\lambda,spot}}{S_{\lambda, phot}}\right)-F_{fac}\left(1-\frac{S_{\lambda, fac}}{S_{\lambda, phot}}\right)} \ ,
\end{equation}
where $F_{spot}$ and $F_{fac}$ are the covering fractions of the spots and faculae and $S_{\lambda, spot}$, $S_{\lambda, fac}$ and $S_{\lambda, phot}$ are the emitted fluxes of the spots, faculae and immaculate photosphere at a given wavelength respectively. An important limitation to consider is that this simplistic method of modelling stellar active regions may be inaccurate, particularly with respect to faculae modelling as it fails to describe any centre-to-limb variations that they may exhibit or their strong dependence on magnetic field strength \citep{norris2017spectral}. However, given the precision of the observations in question this simplified stellar model should be sufficient to begin to investigate any potential heterogeneity displayed by WASP-17. \par
In our retrievals, we considered a varying spot temperature between 4000 and 5000 K, and a faculae temperature between 6600 and 7000 K. Both the spots and faculae coverage fraction bounds where set between 0.0 and 0.9 of the total stellar surface. \par
We tested the possibility of a variable stellar photosphere also on the spectrum in Case 1 and Case 3.

\section{Results}
\subsection{Retrievals accounting for an active star}

\begin{figure*}[htp]
    \centering
    \includegraphics[width=\textwidth]{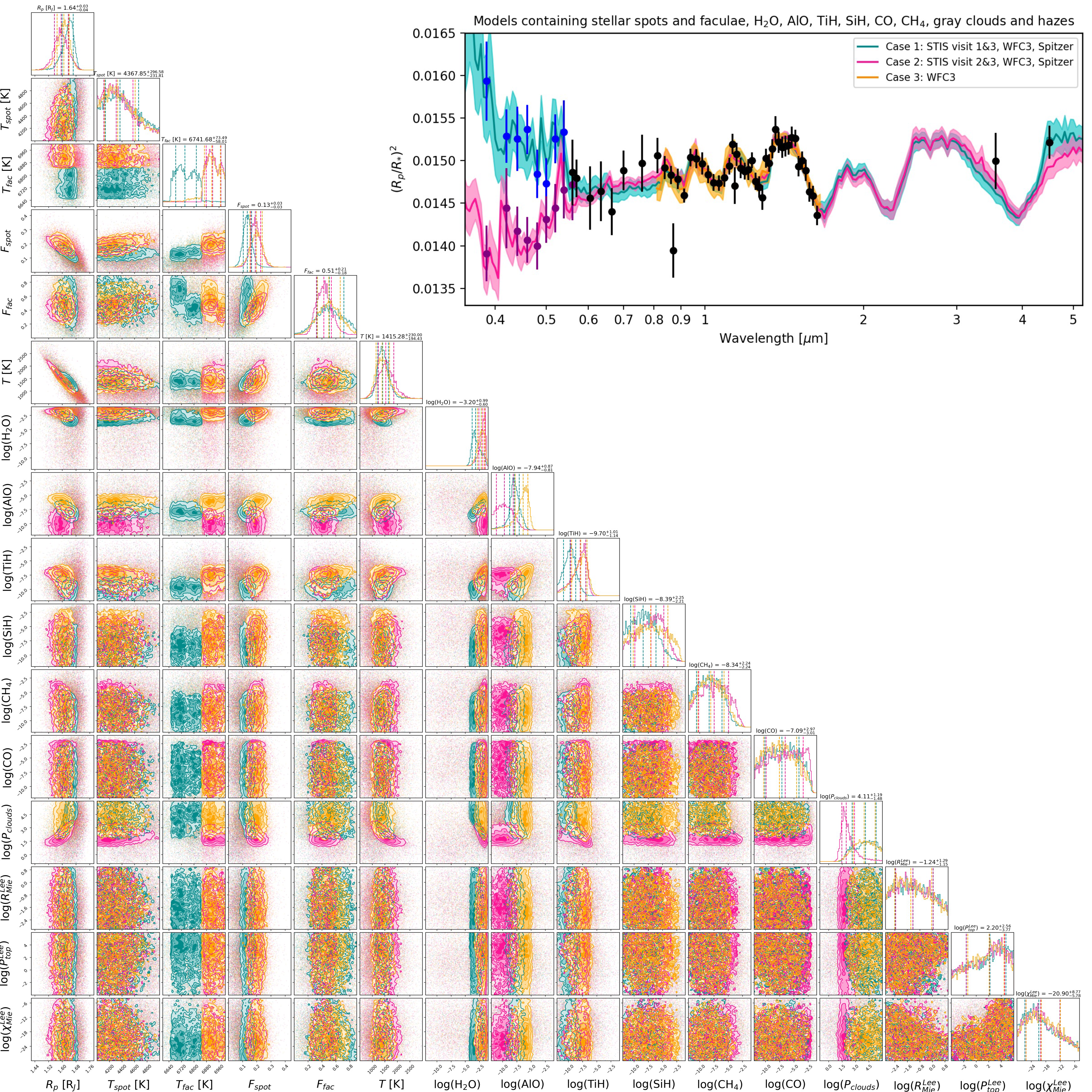}
    \caption{Posterior distributions for the transmission spectrum of WASP-17\,b accounting for an active star retrieved with different datasets. The dashed vertical lines in each histogram refer to the median value, the first quantile (lower bound error) and the third quantile (upper bound error) of each parameter. Similarly, the reported values on top of the histograms denote the parameter's median value, the first and the third quantile. The values shown are those obtained with the spectrum constituted by STIS visit 1\&3, WFC3, Spitzer (Case 1). Inset: transmission spectra with best-fit models and their 1$\sigma$ uncertainties.}
    \label{fig:3posteriors_stellar}
\end{figure*}

\begin{figure*}
    \centering
    \includegraphics[width=\textwidth]{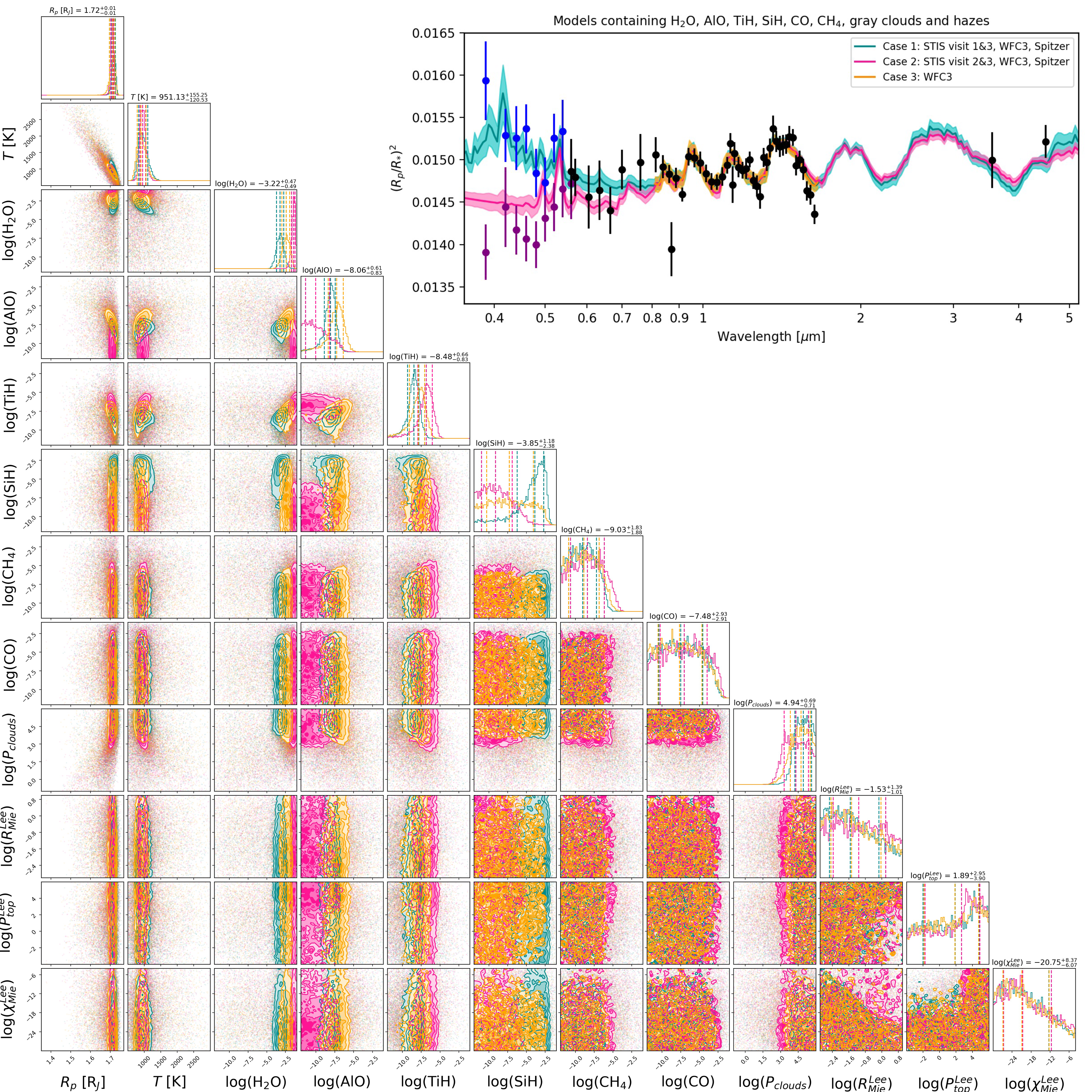}
    \caption{Posterior distributions for the transmission spectrum of WASP-17\,b accounting for a homogeneous star retrieved with different datasets. The dashed vertical lines in each histogram refer to the median value, the first quantile (lower bound error) and the third quantile (upper bound error) of each parameter. Similarly, the reported values on top of the histograms denote the parameter's median value, the first and the third quantile. The values shown are those obtained with the spectrum constituted by STIS visit 1\&3, WFC3, Spitzer (Case 1). Inset: transmission spectra with best-fit models and their 1$\sigma$ uncertainties.}
    \label{fig:3posteriors}
\end{figure*}

When comparing the posteriors for each of the three cases in Figure\,\ref{fig:3posteriors_stellar}, one can find common atmospheric features that are stable independently of the datasets modelled. Firstly, the planetary temperatures retrieved are within 1$\sigma$ of each other and their upper bounds are inside the limits of the equilibrium temperature [1579 K; 1829 K] calculated in Section 2.4. Similarly, the planetary radii at 10 bar are compatible with one another. \par
All three spectra agree on a number of aspects: the presence of water in the solar/super-solar regime which is consistent with the result from \cite{10.1093/mnras/sty2544}, $\log_{10}$(H$_2$O) = -4.04$^{+0.91}_{-0.42}$. The reference solar water abundance, was calculated under thermochemical equilibrium at a pressure of 1 bar for the equilibrium temperature of WASP-17\,b and it is equal to $\log_{10}$(X$^{\odot}_{H_2O}$) = -3.3 \citep{10.1093/mnras/sty2544}. 

Our results suggest the presence of AlO and TiH in the order of 10$^{-8}$ in volume mixing ratio, unconstrained SiH and hazes non-detection. Furthermore the retrievals indicate a strong stellar contamination on the spectrum of WASP-17\,b, in the WFC3 dataset too where we do not expect stellar activity to play a big part. In all three scenarios we retrieved a spot temperature around 4300 K, as well as spots and faculae covering fraction of the star in the order of 10-20\% and 40-50\% respectively. Our results report a faculae temperature in the Case 1 transmission spectrum between 33 and 165 K higher that what Equation\,\ref{eq:faculae} predicts. On the other hand, $T_{fac}$ on Case 2 and Case 3 spectra results between 217 and 322 K higher than expected ($T_{fac}$=6919.28$^{+53.01}_{-40.73}$ and $T_{fac}$=6914.43$^{+55.02}_{-47.08}$). Although the study by \cite{rackham2018transit} does not provide typical uncertainties on the temperature of spots and faculae, we can easily compare the retrieved values with the stellar photospheric uncertainty ($\sim$ 80 K) from Table\,\ref{tab:iraclis_params}. Hence, the $T_{fac}$ retrieved in Case 1, being less than 2$\sigma$ away from the 80 K uncertainty, could be deemed acceptable compared to the faculae temperature in Case 2 and 3, which is three to four times above the 80 K limit.
%the phoenix spectral models only extend up to 7000 K, making the faculae T for the latter two cases a hard upper limit.

In Case 2 and Case 3 spectra we find super-solar water abundances ($\log_{10}$(H$_2$O)=-1.93$_{-0.70}^{+0.39}$ and $\log_{10}$(H$_2$O)=-2.19$_{-0.69}^{+0.55}$ respectively), while Case 1 returns a $\log_{10}$(H$_2$O)=-3.20$_{-0.60}^{+0.99}$, being 0.97 $\times$ solar and 0.79 times the result reported by \cite{barstow2016consistent}, equal to 5$\times$10$^{-4}$. Depending on the spectral scenario considered, the AlO, TiH and SiH abundances retrieved differ (Case 1: $\log_{10}$(AlO)=-7.94$_{-0.81}^{+0.87}$, $\log_{10}$(TiH)=-9.70$_{-1.14}^{+1.01}$, $\log_{10}$(SiH)=-8.39$_{-2.21}^{+2.25}$; Case 2: $\log_{10}$(AlO)=-9.68$_{-1.42}^{+1.55}$, $\log_{10}$(TiH)=-7.91$_{-1.70}^{+0.98}$, $\log_{10}$(SiH)=-7.03$_{-2.84}^{+2.24}$; Case 3: $\log_{10}$(AlO)=-6.39$_{-1.41}^{+0.81}$, $\log_{10}$(TiH)=-7.82$_{-1.59}^{+1.10}$, $\log_{10}$(SiH)=-7.08$_{-3.05}^{+3.02}$;), yet they are compatible with one another being within their 1$\sigma$ confidence band. According to the thermo-chemical equilibrium model GGChem \citep{woitke2018equilibrium}, the expected abundance of AlO and TiH at $p$=1 bar and $T$=$\sim$1400 K is $\sim$10$^{-11}$ and $\sim$10$^{-12}$ in VMR respectively. The theoretical estimate is in accordance with the retrieved AlO abundance in Case 2, where we find a lower bound equal to $\log_{10}$(AlO)=-11.1. However, in all other cases for both AlO and TiH, the retrieved abundances are 3 to 4 sigmas away from the expected values. On the contrary, the retrieved uncertainties on the SiH abundances are in agreement with the theoretical value computed at $p$=1 bar and $T$=$\sim$1400 K, it being $\log_{10}$(SiH)=$\sim$-10.

%The discrepancy could be due to chemical disequilibrium processes or non-solar elemental ratios. These are however processes that we cannot 

%Additionally, the retrieved results are in agreement with the expected solar values for these elements, being $\sim$10$^{-8}$ in volume mixing ratio for AlO and $\sim$10$^{-10}$ for TiH at $p$=1 bar and $T$=1800 K \citep{woitke2018equilibrium}.

We retrieve conflicting results for the cloud deck. The Case 1 and Case 3 posteriors returned a mean cloud deck at a pressure of $\sim$1.1$\times$10$^{4}$ Pa corresponding to $\sim$0.11 bar. In Case 2 the clouds top pressure is higher in the atmosphere, at a pressure of 0.9 mbar. The cloud coverage found at higher altitudes could be caused by the retrieval framework trying to fit the downward slope created by the STIS visit 2 data in the optical range, which cannot be explained by any molecular absorber. The clouds cut off any molecular signature in the optical, leaving just the tip of two particular TiH and SiH signatures visible at 0.54 and 0.42 $\mu$m in the Case 2 spectrum.

Figure\,\ref{fig:scenario1_plus_contributions_stellar} shows Case 1 data fitted by the best-fit model accounting for stellar activity and the contributions of active trace gases, CIA, clouds and hazes to the fit.

We report the retrieval results for each spectral case in Table\,\ref{tab:retrieval} (upper part).

\begin{table*}[htp]
    \centering
        \caption{List of the parameters, their prior bounds, the scale used and the retrieved values (median and standard deviation) for the Case 1, Case 2 and Case 3 spectra assuming an isothermal atmospheric profile.}
    %\begin{tabular}{c@{\hspace{2cm}}c@{\hspace{2cm}}c@{\hspace{2cm}}c@{\hspace{2cm}}c}
    \begin{tabular}{|c|c|c|c|c|c|}
    \multicolumn{6}{c} {Retrieval accounting for an active star}
    \\ 
    \hline \hline
    Retrieved parameters & Prior bounds & Scale & Case 1 & Case 2 & Case 3 \\ 
    & & & STIS 1\&3, WFC3, IRAC & STIS 2\&3, WFC3, IRAC & WFC3 \\ \hline 
    $R_p$ [R$_\text{J}$] & 1 ; 2 & linear & 1.64$_{-0.04}^{+0.03}$ & 1.59$\pm$0.04 & 1.61$_{-0.04}^{+0.03}$\\
    $T_{spot}$ [K] & 4000 ; 5000 & linear & 4367.85$_{-231.81}^{+296.58}$ & 4306.83$_{-196.70}^{+267.22}$ & 4329.70$_{-206.75}^{+278.98}$\\
    $T_{fac}$ [K] & 6600 ; 7000 & linear & 6741.68$_{-58.01}^{+73.49}$ & 6919.28$_{-40.73}^{+53.01}$ & 6914.43$_{-47.08}^{+55.02}$\\
    $F_{spot}$ & 0.0 ; 0.9 & linear & 0.13 $\pm$0.03 & 0.19$_{-0.03}^{+0.04}$ & 0.20$\pm$0.04 \\
    $F_{fac}$ & 0.0 ; 0.9 & linear & 0.51$_{-0.18}^{+0.21}$ & 0.43$_{-0.10}^{+0.11}$ & 0.50$\pm$0.16 \\
    $T$ [K] & 500 ; 3000 & linear & 1415.28$_{-194.43}^{+230}$ & 1539.70$_{-304.42}^{+317.21}$ & 1396.20$_{-243.79}^{+291.34}$ \\
    H$_2$O & -12 ; -1 & $\log_{10}$ & -3.20$_{-0.60}^{+0.99}$ & -1.93$_{-0.70}^{+0.39}$ & -2.19$_{-0.69}^{+0.55}$ \\
    AlO & -12 ; -1 & $\log_{10}$ & -7.94$_{-0.81}^{+0.87}$ & -9.68$_{-1.42}^{+1.55}$ & -6.39$_{-1.41}^{+0.81}$\\
    TiH & -12 ; -1 & $\log_{10}$ & -9.70$_{-1.14}^{+1.01}$ & -7.91$_{-1.70}^{+0.98}$ & -7.82$_{-1.59}^{+1.10}$\\
    SiH & -12 ; -1 & $\log_{10}$ & -8.39$_{-2.21}^{+2.25}$ & -7.03$_{-2.84}^{+2.24}$ & -7.08$_{-3.05}^{+3.02}$\\
    CH$_4$ & -12 ; -1 & $\log_{10}$ & -8.34$\pm$2.24 & -7.48$_{-2.72}^{+2.55}$ & -8.08$_{-2.24}^{+2.34}$\\
    CO & -12 ; -1 & $\log_{10}$ & -7.09$_{-3.01}^{+2.97}$ & -6.51$_{-3.36}^{+3.15}$ & -7.55$_{-2.77}^{+3.03}$\\
    $P_{clouds}$ [Pa] & -1 ; 6 & $\log_{10}$ & 4.11$_{-1.48}^{+1.19}$ & 1.95$_{-0.93}^{+0.48}$ & 3.98$_{-1.03}^{+1.33}$ \\
    $R_{Mie}^{Lee}$ [$\mu$m] & -3 ; 1 & $\log_{10}$ & -1.24$_{-1.15}^{+1.29}$ & -1.25$_{-1.10}^{+1.29}$ & -1.38$_{-1.03}^{+1.33}$\\
    $P_{Mie}^{top}$ [Pa] & -4 ; 6 & $\log_{10}$ & 2.20$_{-3.77}^{+2.56}$ & 2.11$_{-3.71}^{+2.44}$ & 2.08$_{-3.42}^{+2.36}$\\
    $\chi_{Mie}^{Lee}$ & -30 ; -4 & $\log_{10}$ & -20.90$_{-5.78}^{+8.77}$ & -20.13$_{-6.07}^{+8.06}$ & -19.87$_{-6.27}^{+7.91}$ \\ \hline \hline 
    \multicolumn{6}{c} {Retrieval accounting for a homogeneous star} \\ \hline \hline
    Retrieved parameters & Prior bounds & Scale & Case 1 & Case 2 & Case 3 \\ 
    & & & STIS 1\&3, WFC3, IRAC & STIS 2\&3, WFC3, IRAC & WFC3 \\ \hline 
    $R_p$ [R$_\text{J}$] & 1 ; 2 & linear & 1.72$\pm$0.01 & 1.71$\pm$0.01 & 1.71$_{-0.02}^{+0.01}$\\
    $T$ [K] & 500 ; 3000 & linear & 951.13$_{-120.53}^{+155.25}$ & 950.73$_{-83.04}^{+93.36}$ & 900.56$_{-112.25}^{+153.18}$ \\
    H$_2$O & -12 ; -1 & $\log_{10}$ & -3.22$_{-0.49}^{+0.47}$ & -1.40$_{-0.28}^{+0.19}$ & -2.41$_{-0.56}^{+0.48}$ \\
    AlO & -12 ; -1 & $\log_{10}$ & -8.06$_{-0.83}^{+0.61}$ & -10.03$_{-1.35}^{+1.86}$ & -7.25$_{-1.12}^{+0.87}$ \\
    TiH & -12 ; -1 & $\log_{10}$ & -8.48$_{-0.83}^{+0.66}$ & -6.83$_{-1.11}^{+0.79}$ & -7.85$_{-1.22}^{+0.82}$\\
    SiH & -12 ; -1 & $\log_{10}$ & -3.85$_{-2.38}^{+1.18}$ & -9.10$_{-1.87}^{+2.23}$ & -7.26$_{-3.04}^{+3.20}$ \\
    CH$_4$ & -12 ; -1 & $\log_{10}$ & -9.03$_{-1.88}^{+1.83}$ & -8.40$_{-2.26}^{+2.24}$ & -8.87$\pm$2.04 \\
    CO & -12 ; -1 & $\log_{10}$ & -7.48$_{-2.91}^{+2.93}$ & -7.03$_{-3.21}^{+3.08}$ & -7.64$_{-2.87}^{+3.00}$ \\
    $P_{clouds}$ [Pa] & -1 ; 6 & $\log_{10}$ & 4.94$_{-0.71}^{+0.69}$ & 4.31$_{-1.00}^{+1.06}$ & 4.74$_{-0.91}^{+0.80}$ \\
    $R_{Mie}^{Lee}$ [$\mu$m] & -3 ; 1 & $\log_{10}$ & -1.53$_{-1.01}^{+1.39}$ & -1.12$_{-1.24}^{+1.31}$ & -1.47$_{-1.01}^{+1.43}$ \\
    $P_{Mie}^{top}$ [Pa] & -4 ; 6 & $\log_{10}$ & 1.89$_{-3.90}^{+2.95}$ & 2.67$_{-4.42}^{+2.22}$ & 1.87$_{-3.75}^{+2.88}$\\
    $\chi_{Mie}^{Lee}$ & -30 ; -4 & $\log_{10}$ & -20.75$_{-6.07}^{+8.37}$ & -20.63$_{-6.04}^{+9.05}$ & -20.88$_{-6.01}^{+8.56}$\\
    \hline \hline
    \end{tabular}
    \label{tab:retrieval}
\end{table*}

\subsection{Retrievals accounting for a homogeneous star}
We modelled the three spectral scenarios previously identified by accounting for the same molecular and atmospheric parameters. However, in this instance we modelled the star as a homogeneous body. We find that the absence of stellar activity does not impact the molecular detections, but it does affect a few of the retrieved abundances. 

As in the previous case, our data suggests the presence of water, aluminium oxide and titanium hydride. Our retrievals return solar $\log_{10}$(H$_2$O)=-3.22$_{-0.49}^{+0.47}$ and super-solar $\log_{10}$(H$_2$O)=-2.41$_{-0.56}^{+0.48}$, $\log_{10}$(H$_2$O)=-1.40$_{-0.28}^{+0.19}$ water concentrations when modelling the Case 1, Case 3 and Case 2 spectrum respectively. The latter two results indicate a higher abundance than what current theories predict for hot-Jupiters ($\log_{10}$(H$_2$O)=$\sim$-3.3) \citep{venot2012chemical, venot2015chemical, woitke2018equilibrium}.

The AlO and TiH abundances retrieved from the Case 3 spectrum appear to be more compatible to those retrieved from Case 1. However, both the AlO and the TiH results obtained from each of the three cases are within their respective 1$\sigma$ uncertainties. These results are not consistent with chemical equilibrium models, as \cite{woitke2018equilibrium} predict a volume mixing ratio of roughly 10$^{-16}$ for AlO and 10$^{-18}$ for TiH at a pressure of 1 bar and $T$=$\sim$900 K. The discrepancies could be caused by chemical disequilibrium processes, e.g. photochemistry \citep{venot2015new, fleury2019photochemistry} or quenching \citep{shulyak2020stellar}, occurring in the atmosphere of WASP-17\,b.

SiH is still largely undetected in Case 2 and Case 3 spectrum. On the other hand, exceptionally high amounts of this trace gas are detected in Case 1. We recognise that such large contribution ($\log_{10}$(SiH)=-3.85$_{-2.38}^{+1.18}$) is unlikely as theoretical models \citep{woitke2018equilibrium} predict a SiH volume mixing ratio of $\sim$10$^{-16}$ at 1 bar and at $\sim$900 K. Moreover, all three spectra return a consistent median planetary radius, temperature and cloud coverage, approximately equal to 1.71 R$_\text{J}$, 930 K and $\sim$4.6$\times$10$^{4}$ Pa respectively.

In place of stellar faculae, a Mie scattering haze could have explained the upward slope in Case 1 spectrum. Unfortunately, this option is refused by the retrieval, which is not able to constrain any of the aerosol parameters: particle size $R_{Mie}^{Lee}$, top pressure $P_{Mie}^{top}$, mixing ratio $\chi_{Mie}^{Lee}$. Even when we test the aerosols presence in the simplest model containing only water and clouds, their posterior distributions are completely unconstrained. 

The retrieved values for each atmospheric parameter are reported in Table\,\ref{tab:retrieval} (bottom part) while Figure\,\ref{fig:scenario2_plus_contributions_no_stellar} shows Case 2 data fitted by the best-fit model and the contributions of active trace gases, CIA, clouds and hazes to the fit.

\section{Discussion}
% What experiments did you conduct and what were the results?
% What do the results mean?
% What were the important results from your study?
% How did the results answer your research questions?
% Did your results support your hypothesis or reject your hypothesis?
% What are the variables or factors that might affect your results?
% What were the strengths and limitations of your study?
% What other published works support your findings?
% What other published works contradict your findings?
% What possible factors might cause your findings different from other findings?
% What is the significance of your research?
% What are new research questions to explore based on your findings?

\begin{table*}[htp]
    \centering
        \caption{Bayesian log-difference between different models depending on the observations fed to the retrievals.}
    %\begin{tabular}{c@{\hspace{0.2cm}}c@{\hspace{0.2cm}}c@{\hspace{0.2cm}}c}
    \begin{tabular}{cccc}
    \hline \hline
    \multicolumn{4}{c} {Case 1: STIS visit 1\&3, WFC3, Spitzer} \\ \hline \hline
    Setup & Log Evidence & Sigma & Retrieved Temperature \ [K] \\
    & & (discarded by) \\ \hline 
    (1) H$_2$O, AlO, TiH, SiH, CH$_4$, CO, clouds, hazes, stellar activity & 395.73 & & 1415.28$_{-194.43}^{+230.00}$ \\
    (2) H$_2$O, AlO, TiH, SiH, CH$_4$, CO, clouds, hazes & 389.31 & 4.01 w.r.t. (1) & 951.13$_{-120.53}^{+155.25}$ \\
    (3) H$_2$O, clouds, stellar activity & 395.02 & 1.84 w.r.t. (1) & 1429.50$_{-245.74}^{+363.20}$\\
    (4) H$_2$O, AlO, TiH, SiH, clouds, hazes, stellar activity & 395.90 & $<$ 1 w.r.t. (1) & 1397.57 $_{-193.95}^{+229.83}$ \\
    (5) H$_2$O, TiO, VO, FeH, clouds, stellar activity & 393.60 & 2.6 w.r.t. (1) & 1854.96$_{-495.03}^{+382.33}$ \\
    \hline \hline \\

    \multicolumn{4}{c} {Case 2: STIS visit 2\&3, WFC3, Spitzer} \\ \hline \hline
    Setup & Log Evidence & Sigma & Retrieved Temperature \ [K] \\
    & & (discarded by) \\ \hline 
    (1) H$_2$O, AlO, TiH, SiH, CH$_4$, CO, clouds, hazes, stellar activity & 401.41 & & 1539.70$_{-304.42}^{+317.21}$\\
    (2) H$_2$O, AlO, TiH, SiH, CH$_4$, CO, clouds, hazes & 387.38 & 5.64 w.r.t. (1) & 950.73$_{-83.04}^{+93.36}$ \\
    (3) H$_2$O, clouds, stellar activity & 401.68 & $<$ 1 w.r.t. (1) & 1357.63$_{-263.14}^{+234.71}$\\
    (4) H$_2$O, AlO, TiH, SiH, clouds, hazes, stellar activity & 401.14 & $<$ 1 w.r.t. (1) & 1435.55$_{-282.96}^{+266.26}$ \\
    (5) H$_2$O, TiO, VO, FeH, clouds, stellar activity & 401.42 & $<$ 1 w.r.t. (1) & 1550.71$_{-346.10}^{+306.26}$ \\
    \hline \hline \\
    
    \multicolumn{4}{c} {Case 3: WFC3} \\ \hline \hline
    Setup & Log Evidence & Sigma & Retrieved Temperature \ [K] \\ 
    & & (discarded by) \\ \hline
    (1) H$_2$O, AlO, TiH, SiH, CH$_4$, CO, clouds, hazes, stellar activity & 285.37 & & 1396.20$_{-243.79}^{+291.34}$\\
    (2) H$_2$O, AlO, TiH, SiH, CH$_4$, CO, clouds, hazes & 278.77 & 4.05 w.r.t. (1) & 900.56$_{-112.25}^{+153.18}$\\
    (3) H$_2$O, clouds, stellar activity & 284.59 & 1.88 w.r.t. (1) & 1374.81$_{-262.77}^{+378.09}$\\
    (4) H$_2$O, AlO, TiH, SiH, clouds, hazes, stellar activity & 285.44 & $<$ 1 w.r.t. (1) & 1356.01$_{-220.37}^{+256.60}$\\
    (5) H$_2$O, TiO, VO, FeH, clouds, stellar activity & 284.15 & 2.15 w.r.t. (1) & 1475.32$_{-311.57}^{+309.41}$ \\
    \hline \hline \\
    \end{tabular}
    \label{tab:sigma_evidence}
\end{table*}

WASP-17\,b has been observed, studied and modelled intensively in the past, e.g. \cite{anderson2011thermal, mandell2013exoplanet, 10.1093/mnras/sty2544, goyal2020library}, by employing observations taken with WFC3/G141
(staring mode), HST/STIS and
Spitzer. Our study, complementary to the work by \cite{alderson2022comprehensive}, offers a consistent analysis performed by combining and reducing in a reproducible manner all currently available WASP-17\,b transmission data obtained with space-based observatories. We include all public STIS and IRAC data, plus WFC3/G141 and WFC3/G102 observations both taken in spatial scanning mode. This large breadth of observations allows us to investigate the planet's atmospheric spectrum spanning the wavelengths from the optical to the near infrared.

The combination of several datasets has its pros and cons. By coupling observations taken at different wavelengths, we are able to achieve a broad wavelength coverage, paramount to obtain more precise measurements and unambiguous molecular detection. It is extremely challenging to absolutely calibrate data obtained from different detectors, especially if each of them targets a very specific spectral range with a limited or even absent wavelength overlap \citep{yip2020compatibility}. Moreover, each instrument creates its own distinct systematics that need to be modelled ad hoc. 

The STIS detector shares a short portion of the wavelength space, about 0.14 $\mu$m, with WFC3. This wavelength overlap is useful to investigate potential calibration discrepancies between instruments. In the 0.8-0.94 $\mu$m range, our spectrum displays six WFC3/G102 datapoints and one STIS/G750L point (Fig.\,\ref{fig:complete_spectrum}). The  response of the STIS G750L grating shows a strong dip above  0.8 $\mu$m \citep{prichard2022stis}, hence the last spectral bin (0.8- 0.94 $\mu$m) has a flux rate $\sim$50\% lower than the rest of the bands and it is possibly subject to different systematics. Because of the shape of the transmission curve, we opted to increase the size of the final spectral bin to counteract the flux loss. Our analysis returns a STIS data point at 0.87 $\mu$m that is at least 2$\sigma$ away from the G102 datapoints placed between 0.8 and 0.94 $\mu$m. A similar decrease in transit depth at the same wavelengths was also recovered by \cite{sing2016continuum}, as it can be seen in Figure\,\ref{fig:complete_spectrum}. Yet, the transit depths of the white lightcurves for G750L and G102 are within 1$\sigma$ of with each other (WFC3/G102: $R_p/R_*$=0.12184$_{-0.00022}^{+0.00019}$; STIS/G750L: $R_p/R_*$=0.1211$_{-0.0013}^{+0.0014}$), confirming that the two instrumental calibrations are in overall agreement. Finally, we must note that this problematic STIS/G750L datapoint is not statistically significant for the spectral modelling. From Figure\,\ref{fig:3posteriors_stellar} and \ref{fig:3posteriors} we can observe that the TauREx\,3 atmospheric model ignores this data point when computing the best-fit spectrum.

On the other hand, between 3 and 5 $\mu$m lie the Spitzer data points, whose transit depths cannot be compared to any additional dataset. We shifted the two Spitzer data points 100 ppm up and 100 ppm down from their original position to check if the results would be considerably impacted. In either case the retrieved mean molecular abundances and other atmospheric features such as radius and temperature, fluctuate at the order of the fifth decimal figure from the initial values, implying an accurate data reduction process and stability of the results. \par
An additional issue to take into account when combining datasets taken months or even years apart is atmospheric variability. We must remember that planets are dynamic bodies, varying in space and time. Hence, mechanisms like star-planet interaction and global circulation processes such as polar vortexes and zonal jets can produce very strong temperature and chemical differences \citep{cho2003changing, cho2021storms}. It might not be surprising that the same planet can display contrasting spectral features at the same wavelengths, such as the two opposite spectral trends we presented between $\sim$0.4 and $\sim$0.6 $\mu$m. In light of this, we must stress that the publicly available STIS observations used in this study, lack the post-egress part of the planetary transit (Figure\,\ref{fig:transit_curve_stis1}, \ref{fig:transit_curve_stis2}, \ref{fig:transit_curve_stis3}). Missing lightcurve data strongly impacts the transit fit model, which then could potentially affect the derived transit depth, leading to discordant spectral features amongst different observations, as highlighted in Figure\,\ref{fig:complete_spectrum}. \par
Due to the discrepancies in the data, we decided to model the contrasting STIS datasets separately: in Case 1 we considered a spectrum that includes STIS visit 1\&3, WFC3, IRAC; in Case 2 we modelled the spectrum with STIS visit 2\&3, WFC3, IRAC. Additionally, we employed the WFC3 observations alone (Case 3) to check how the results would change when we do not include the problematic STIS datasets. \par
%On top of that, in the results section we presented how a model containing either a heterogeneous or a homogeneous star could be employed to describe the spectral scenarios we identified. 
To explain the unusually strong downward slope in the optical range displayed by the Case 2 spectrum (Figure\,\ref{fig:3posteriors_stellar}, inset), an active star was included in the retrieval set-up. For consistency, we applied the same stellar model to Case 1 and Case 3 spectra. Potentially, STIS visit 2\&3, being taken just a few days apart, were impacted by a stellar event, causing the spectrum to be best fitted by a model containing a high spots and faculae covering fraction. STIS visit 1 could have also been affected by a similar event, given that its best model favours stellar activity in place of aerosols. \par
Theoretically, a faculae covering fraction between 40 and 50\% is unlikely in a F6-type star like WASP-17, unless we are dealing with a rare and extreme event. Moreover, we recognise that the posterior distributions between the faculae temperature and their covering fraction is degenerate (see Figure\,\ref{fig:3posteriors_stellar}). This is because the retrieval is unable to distinguish between two scenarios: 1) a smaller covering fraction of hotter faculae and 2) a larger covering fraction of cooler faculae. Both scenarios result in similar stellar disk-integrated SEDs and thus introduce similar contamination features to the transmission spectrum. \par
There are no previous studies on the activity of the star so we are unable to compare our results with published research. However, \cite{2018A&A...618A..98K} mention that their attempt to identify a sodium feature in the spectrum of WASP-17\,b could have been impacted by an in-transit activity of the host star, which decreased the planetary Na signal. Hence, stellar contamination cannot be excluded as an additional cause for a decreasing transit depth in the optical regime. Continuous photometric monitoring of WASP-17 is necessary to determine accurately the activity levels of the star. \par
When the star is treated as an homogeneous body, the retrieval of Case 1 spectrum results in a extremely high SiH volume mixing ratio $\log_{10}$(SiH)=-3.85$_{-2.38}^{+1.18}$. This result is unlikely, given that chemical equilibrium models predict a SiH abundance in the order of 10$^{-16}$ at 900 K \citep{woitke2018equilibrium}. Perhaps, such high SiH abundance is due to a missing molecule which is not accounted for. Either this missing gas absorbs at the same wavelengths of SiH or it has similar absorption features. Even if we wanted to investigate this further, we do not possess all the possible molecular cross-sections and it is not feasible to test all the known molecules that absorb in the optical region. However, by inspecting the SiH results retrieved from the Case 2 and Case 3 spectra, we confirm that SiH is largely unconstrained and remains undetected. 
\par
Overall, we notice that when we consider an inactive star, the retrievals struggle to fit the data points at the end of the WFC3 wavelength coverage at $\sim$ 1.6 - 1.8 $\mu$m and those at 0.4 - 0.55 $\mu$m, especially visit 2 (see Figure\,\ref{fig:3posteriors}, inset). This statement is supported by the log evidence of the models: when stellar activity is included, the log(E) increases by a minimum of 4$\sigma$ (compared to the same model without stellar activity). Furthermore, independently of the datasets modelled, our results suggest the presence of the same trace gases (H$_2$O, AlO, TiH) as when accounting for a heterogeneous star. %The abundances retrieved vary from spectrum to spectrum, hence it is possible that the true parameter values lie somewhere in between the results retrieved from each scenario. However, the AlO and TiH VMRs retrieved are in agreement with expected abundances from equilibrium chemistry models.} 
At this stage we do not possess enough evidence to discard a particular STIS dataset in favour of another one, but the information content derived by WFC3 data alone is enough to propose the existence of those species in the atmosphere of WASP-17\,b. \par
No previous study has detected any metal hydrates or oxides in the atmosphere of this particular hot-Jupiter, albeit \cite{bento2014optical} searching for TiO signatures in the upper atmosphere of WASP-17\,b. Theoretical studies by \cite{fortney2008unified} propose the presence of TiO and VO in the atmosphere of planets with large day-night temperature contrasts. We do not exclude the presence of these trace gases but our analyses favour a model containing AlO, TiH, SiH, CH$_4$ and CO rather than TiO, VO and FeH by a 2.6 $\sigma$ evidence, when applied to the spectrum in Case 1 (Table\,\ref{tab:sigma_evidence}). As of yet, this is the first time that the presence of AlO and TiH has been suggested in the atmosphere of WASP-17\,b. \par 
During our preliminary retrievals, we tested different models containing a variety of trace gases on the spectrum in Case 1, but ultimately we found that including H$_2$O, AlO, TiH, SiH, clouds, hazes and an active star would lead to the best results. Adding CH$_4$ and CO in the latter model did not substantially affect the Bayesian evidence, the two models being 0.26$\sigma$ away from each other. 

On the other hand, the spectrum containing STIS visit 2 proved trickier to model. Out of the five models tested on these data, model (1), (3), (4) and (5) in Table\,\ref{tab:sigma_evidence} (middle panel) were within less than 1$\sigma$ of each other, and model (2) containing H$_2$O, AlO, TiH, SiH, CH$_4$, CO clouds and hazes was 5.64$\sigma$ away from model (1), which includes the same chemical species plus stellar variability. Ultimately, we used the best-fit model for the Case 1 spectral scenario also in the Case 2 dataset and on the WFC3 data alone, to make a valid comparison between the scenarios in terms of molecular abundances. The retrievals conducted exclusively on the WFC3 data too favour a model containing H$_2$O, AlO, TiH, SiH, clouds, hazes, stellar spots and faculae. 

Our results suggest that WASP-17\,b possesses mostly a clear atmosphere, free from a thick layer of clouds and aerosols. \cite{sing2016continuum} report the atmosphere of this hot-Jupiter to be the clearest among the 10 exoplanets in their study, presenting distinct water features. \cite{10.1093/mnras/sty2544} find similar results, their spectrum being fitted best by a model containing mostly a cloudless atmosphere with a cloud/haze fraction of 0.2 and a cloud top pressure of 0.1 mbar. On the other hand, \cite{barstow2016consistent} report the presence of scattering aerosols at relatively high altitudes, but not so strong to reduce the size of the other molecular features. \cite{mandell2013exoplanet} also claim that their data is best fitted by a model that accounts for hazes or clouds. \par  
We were unable to identify the alkali lines (Na, K) reported by \cite{sing2016continuum}. Their data point at $\sim$0.6 $\mu$m has an uncertainty three times larger than the other ones and it is not statistically significant to demonstrate the presence of sodium in the atmosphere of WASP-17\,b.

Furthermore, during our exploratory retrievals we did not find evidence of CO$_2$ which, on the contrary, is detected by \cite{alderson2022comprehensive}. Due to the sparse nature of the Spitzer datapoints and the influence that additional molecules can impart in the spectrum of an exoplanet at the wavelengths probed by IRAC, it is challenging to place precise constraints on the abundance of carbon-bearing molecules.

Regardless of whether we consider an active star or not, the retrieved planetary temperature is always lower than the calculated equilibrium temperature, as transmission studies often find. Discrepancies between the retrieved and the calculated temperatures have been analysed by several authors, e.g. \cite{caldas2019effects, 2020ApJ...893L..43M, pluriel2020strong, changeat2021exploration}, and found to arise from a combination of aspects: the adoption of 1D atmospheric models to describe the different molecular content of the morning and evening terminators, the use of simplified models to characterise the planets' 3D processes and the adoption of incorrect assumptions regarding the albedo and the emissivity when calculating the planet's equilibrium temperature. Hence, retrieving a lower temperature than calculated should not be surprising.

\section{Conclusion}
Remote-sensing studies are the only technique through which we can unveil the atmospheric characteristics of planets outside our solar system. In recent years the \textit{Hubble Space Telescope} has pioneered the spectroscopic investigation of exoplanets. In this study we presented the transmission spectrum of WASP-17\,b observed with the Space Telescope Imaging Spectrograph and the Wide Field Camera 3 mounted on HST, and the Infrared Camera Array aboard the Spitzer Space Telescope. The Hubble spectroscopic data was reduced with Iraclis, a specialised STIS and WFC3 data reduction routine. Photometric data obtained by Spitzer was instead reduced and analysed with the TLCD-LSTM method by means of a long short-term memory network. Given the discordant results obtained for two STIS observations which cover the same wavelength range, we decided to model the spectrum of WASP-17\,b in three separate cases: (a) including STIS visit 1\&3, WFC3, IRAC (Case 1); (b) including STIS visit 2\&3, WFC3, IRAC (Case 2); (c) WFC3 only (Case 3). The fully Bayesian retrieval framework TauREx\,3 was then employed to find the model that can best explain each spectral scenario. Retrieval results find that the best-fit model is constituted by H$_2$O, AlO, TiH, SiH, clouds, hazes and stellar activity in Case 1 and 3. The best-fit model for Case 2 includes CH$_4$ and CO too, however their addition only improved the Bayesian evidence by 0.26$\sigma$. We acknowledge that the extreme stellar activity identified on the F6-type host star is unlikely, therefore we check if a model containing the same trace gases but including an homogeneous star is able to describe all three spectra. Independently of the activity of the star, our analysis indicates the presence of water in the solar/super-solar regime and possible traces of AlO and TiH in the atmosphere of WASP-17\,b. The atmosphere of the inflated hot-Jupiter appears to be free from both an optically thick layer of gray clouds and hazes. Given the incompleteness of the STIS lightcurves, we reckon that additional HST/STIS observations will help to improve our knowledge about the limb of this gas giant. In fact, the remarkable large scale height of WASP-17\,b, makes the planet a perfect target for further atmospheric characterisation, both from the ground and by current and next generation space-based observatories such as JWST \citep{greene2016characterizing} and Ariel \citep{tinetti2018chemical}.

\section*{Acknowledgements}
We thank our anonymous referee for the insightful comments which have improved the quality of our work.
\textit{Data}: This work is based upon observations with the NASA/ESA Hubble Space Telescope, obtained at the Space Telescope Science Institute (STScI) operated by AURA, Inc. The publicly available HST observations presented here were taken as part of proposal 12473 and 14918, led by David Sing and Hannah Wakeford respectively. These were obtained from the Hubble Archive which is part of the Mikulski Archive for Space Telescopes. This work is also based in part on observations made with the Spitzer Space Telescope, which is operated by the Jet Propulsion Laboratory, California Institute of Technology, under a contract with NASA. The publicly available Spitzer data were taken as part of programme 90092 led by Jean-Michel Desert. This paper also uses data collected by the TESS mission, which are publicly available from the Mikulski Archive for Space Telescopes (MAST) and produced by the Science Processing Operations Center (SPOC) at NASA Ames Research Center \citep{jenkins2016tess}. %This research ef- fort made use of systematic error-corrected (PDC-SAP) photometry (Smith et al. 2012; Stumpe et al. 2012, 2014). 
Funding for the TESS mission is provided by NASAs Science Mission directorate. We are thankful to those who operate this archive, the public nature of which increases scientific productivity and accessibility \citep{peek2019robust}.

\textit{Funding}:
This project has received funding from the European Research Council under the European Union’s Horizon 2020 research and innovation program (grant agreement No. 758892, ExoAI) and the European Union’s Horizon 2020 COMPET program (grant agreement No. 776403, ExoplANETS A). Furthermore, we acknowledge funding by the UK Space Agency and Science and Technology Funding Council grants ST/K502406/1, ST/P000282/1, ST/P002153/1, ST/ S002634/1, ST/T001836/1, ST/V003380/1 and ST/W00254X/1. BE is a Laureate of the Paris Region fellowship programme which is supported by the Ile-de-France Region and has received funding under the Horizon 2020 innovation framework programme and the Marie Sklodowska-Curie grant agreement no. 945298.

\textit{Software}: Iraclis \citep{tsiaras2016detection}, TauREx\,3 \citep{Al_Refaie_2021}, PyLightcurve \citep{tsiaras2016new}, Lightkurve \citep{cardoso2018lightkurve}, ExoTETHyS \citep{morello2020exotethys}, Astropy \citep{price2018astropy}, h5py \citep{collette2013python}, emcee \citep{foreman2013emcee}, Matplotlib \citep{hunter2007matplotlib}, Multinest \citep{feroz2009multinest}, PyMultinest \citep{buchner2014x}, Pandas \citep{mckinney2011pandas}, Numpy \citep{oliphant2006guide}, SciPy \citep{virtanen2020scipy}, corner \citep{foreman2016corner}.

\clearpage
\bibliographystyle{aasjournal}
\bibliography{main}

\clearpage
\appendix
\section{Network parameters used for detrending IRAC observations}

\begin{table*}[htp]
\centering
\caption{Table of parameters used for detrending IRAC observations of WASP-17\,b.}
\begin{tabular}{|l|l|l|}
\hline
                             & Irac Channel 1 & Irac Channel 2 \\ \hline
Program                      & $90092$          & $90092$          \\ \hline
Aorkey                       & $47040000$       & $47039488$       \\ \hline
Radius [pixels]              & $3.25$           & $3.25$           \\ \hline
Centroiding method           & 2D Gaussian    & 2D Gaussian    \\ \hline
Start of test set [BMJD TDB] & $56422.595$      & $56426.330$      \\ \hline
End of test set [BMJD TDB]   & $56422.783$      & $56426.518$      \\ \hline
\ \# layers                    & $2$              & $2$              \\ \hline
\ \# units                     & $512$            & $512$            \\ \hline
Dropout rate                 & $10\%$           & $10\%$           \\ \hline
Adam learning rate           & $0.005$          & $0.005$          \\ \hline
$\beta$ decay rate           & $0.95$           & $0.95$           \\ \hline
\end{tabular}
\label{tab:spitzer_network_params}
\end{table*}

\clearpage

\section{Spectral data}
\begin{table*}[htp]
    \centering
        \caption{Reduced and fitted spectral data from the raw HST/STIS transmission data using Iraclis.}
    %\begin{tabular}{c@{\hspace{1.5cm}}c@{\hspace{1.5cm}}c@{\hspace{1.5cm}}c@{\hspace{1.5cm}}c}
    \begin{tabular}{cccccc}
    \multicolumn{6}{c} {} \\ \hline \hline
    Wavelength [$\mu$m] & Transit depth [\%] & Error [\%] & Bandwidth [$\mu$m] & Instrument & Grating\\ \hline 
    0.385 & 1.593 & 0.045 & 0.050 & HST STIS & G430L \\
    0.420 & 1.529 & 0.030 & 0.020 & HST STIS & G430L \\
    0.440 & 1.525 & 0.037 & 0.020 & HST STIS & G430L \\
    0.460 & 1.537 & 0.028 & 0.020 & HST STIS & G430L \\
    0.480 & 1.484 & 0.027 & 0.020 & HST STIS & G430L \\
    0.500 & 1.473 & 0.029 & 0.020 & HST STIS & G430L \\
    0.520 & 1.525 & 0.027 & 0.020 & HST STIS & G430L \\
    0.540 & 1.534 & 0.036 & 0.020 & HST STIS & G430L \\
    0.560 & 1.472 & 0.041 & 0.020 & HST STIS & G430L \\ 
    \\
    0.385 & 1.391 & 0.032 & 0.050 & HST STIS & G430L \\ 
    0.420 & 1.444 & 0.046 & 0.020 & HST STIS & G430L \\ 
    0.440 & 1.417 & 0.028 & 0.020 & HST STIS & G430L \\ 
    0.460 & 1.407 & 0.026 & 0.020 & HST STIS & G430L \\ 
    0.480 & 1.400 & 0.027 & 0.020 & HST STIS & G430L \\ 
    0.500 & 1.431 & 0.026 & 0.020 & HST STIS & G430L \\ 
    0.520 & 1.444 & 0.027 & 0.020 & HST STIS & G430L \\ 
    0.540 & 1.466 & 0.033 & 0.020 & HST STIS & G430L \\ 
    0.560 & 1.486 & 0.037 & 0.020 & HST STIS & G430L \\ 
    \\
    0.570 & 1.479 & 0.021 & 0.040 & HST STIS & G750L \\ 
    0.605 & 1.456 & 0.036 & 0.030 & HST STIS & G750L \\ 
    0.635 & 1.464 & 0.034 & 0.030 & HST STIS & G750L \\ 
    0.665 & 1.440 & 0.027 & 0.030 & HST STIS & G750L \\ 
    0.700 & 1.488 & 0.022 & 0.040 & HST STIS & G750L \\ 
    0.760 & 1.497 & 0.033 & 0.080 & HST STIS & G750L \\ 
    0.870 & 1.394 & 0.031 & 0.140 & HST STIS & G750L \\ 
    \hline
    \end{tabular}
    \label{tab:stis_data}
\end{table*}  
   
\begin{table}[htp]
    \centering
        \caption{Reduced and fitted spectral data from the raw Spitzer/IRAC transmission data using TLCD-LSTM.}
    \begin{tabular}{ccccc}
    \multicolumn{5}{c} {} \\ \hline \hline
    Wavelength [$\mu$m] & Transit depth [\%] & Error [\%] & Bandwidth [$\mu$m] & Instrument\\ \hline 
    3.560 & 1.500 & 0.032 & 0.380 & Spitzer IRAC \\
    4.500 & 1.521 & 0.018 & 0.560 & Spitzer IRAC \\
    \hline
    \end{tabular}
    \label{tab:spitzer_data}
\end{table}

\begin{table}[htp]
    \centering
    \caption{Reduced and fitted spectral data from the raw HST/WFC3 transmission data using Iraclis.}
    \begin{tabular}{cccccc}
    \multicolumn{6}{c} {} \\ \hline \hline
    Wavelength [$\mu$m] & Transit depth [\%] & Error [\%] & Bandwidth [$\mu$m] & Instrument & Grism\\ \hline 
    0.813 & 1.506 & 0.019 & 0.025 & HST WFC3 & G102 \\ 
    0.838 & 1.492 & 0.012 & 0.025 & HST WFC3 & G102 \\ 
    0.863 & 1.483 & 0.009 & 0.025 & HST WFC3 & G102 \\ 
    0.888 & 1.479 & 0.011 & 0.025 & HST WFC3 & G102 \\ 
    0.913 & 1.459 & 0.007 & 0.025 & HST WFC3 & G102 \\ 
    0.938 & 1.504 & 0.009 & 0.025 & HST WFC3 & G102 \\ 
    0.963 & 1.502 & 0.010 & 0.025 & HST WFC3 & G102 \\ 
    0.988 & 1.497 & 0.012 & 0.025 & HST WFC3 & G102 \\ 
    1.013 & 1.484 & 0.009 & 0.025 & HST WFC3 & G102 \\ 
    1.038 & 1.474 & 0.009 & 0.025 & HST WFC3 & G102 \\ 
    1.063 & 1.474 & 0.008 & 0.025 & HST WFC3 & G102 \\ 
    1.088 & 1.480 & 0.010 & 0.025 & HST WFC3 & G102 \\ 
    1.113 & 1.493 & 0.009 & 0.025 & HST WFC3 & G102 \\ 
    1.138 & 1.471 & 0.020 & 0.025 & HST WFC3 & G102 \\ 
    \\
    1.126 & 1.519 & 0.011 & 0.022 & HST WFC3 & G141 \\
    1.148 & 1.507 & 0.012 & 0.021 & HST WFC3 & G141 \\
    1.169 & 1.491 & 0.011 & 0.021 & HST WFC3 & G141 \\
    1.189 & 1.488 & 0.009 & 0.020 & HST WFC3 & G141 \\
    1.208 & 1.489 & 0.011 & 0.019 & HST WFC3 & G141 \\
    1.228 & 1.500 & 0.008 & 0.019 & HST WFC3 & G141 \\
    1.246 & 1.476 & 0.009 & 0.019 & HST WFC3 & G141 \\
    1.266 & 1.469 & 0.011 & 0.019 & HST WFC3 & G141 \\
    1.285 & 1.457 & 0.013 & 0.019 & HST WFC3 & G141 \\
    1.304 & 1.503 & 0.010 & 0.019 & HST WFC3 & G141 \\
    1.323 & 1.497 & 0.009 & 0.019 & HST WFC3 & G141 \\
    1.341 & 1.514 & 0.010 & 0.019 & HST WFC3 & G141 \\
    1.361 & 1.537 & 0.014 & 0.019 & HST WFC3 & G141 \\
    1.380 & 1.522 & 0.012 & 0.020 & HST WFC3 & G141 \\
    1.400 & 1.516 & 0.011 & 0.020 & HST WFC3 & G141 \\
    1.420 & 1.517 & 0.011 & 0.020 & HST WFC3 & G141 \\
    1.441 & 1.518 & 0.011 & 0.021 & HST WFC3 & G141 \\
    1.462 & 1.527 & 0.012 & 0.021 & HST WFC3 & G141 \\
    1.483 & 1.526 & 0.009 & 0.022 & HST WFC3 & G141 \\
    1.505 & 1.493 & 0.011 & 0.022 & HST WFC3 & G141 \\
    1.528 & 1.500 & 0.009 & 0.023 & HST WFC3 & G141 \\
    1.552 & 1.488 & 0.011 & 0.024 & HST WFC3 & G141 \\
    1.576 & 1.463 & 0.010 & 0.025 & HST WFC3 & G141 \\
    1.602 & 1.459 & 0.011 & 0.026 & HST WFC3 & G141 \\
    1.629 & 1.436 & 0.010 & 0.028 & HST WFC3 & G141 \\
    \hline
    \end{tabular}
    \label{tab:wfc3_data}
\end{table}

\clearpage
\section{Retrieval contribution functions}
\begin{figure*}[htp]
    \centering
    \includegraphics[width=\textwidth]{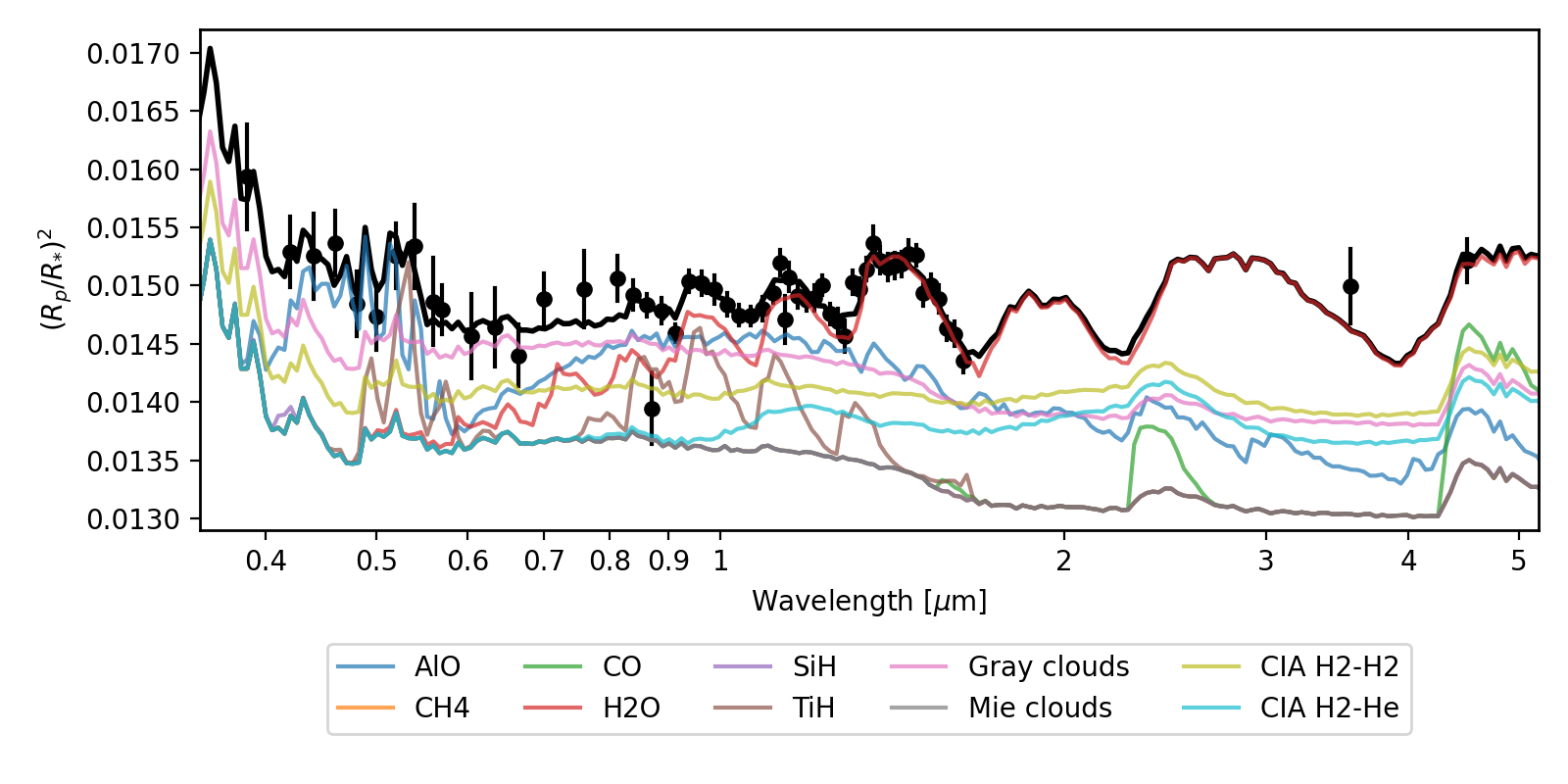}
    \caption{STIS visit 1\&3, WFC3 and Spitzer data (Case 1) fitted by the best-fit model (black line) that includes an active star together with the contributions from CIA, active trace gases, clouds and hazes. Stellar activity induces wavelength variations at the base of the model, reason why the gray clouds and Mie clouds contributions are not flat at the bottom.}
    \label{fig:scenario1_plus_contributions_stellar}
\end{figure*}

\begin{figure*}[htp]
    \centering
    \includegraphics[width=\textwidth]{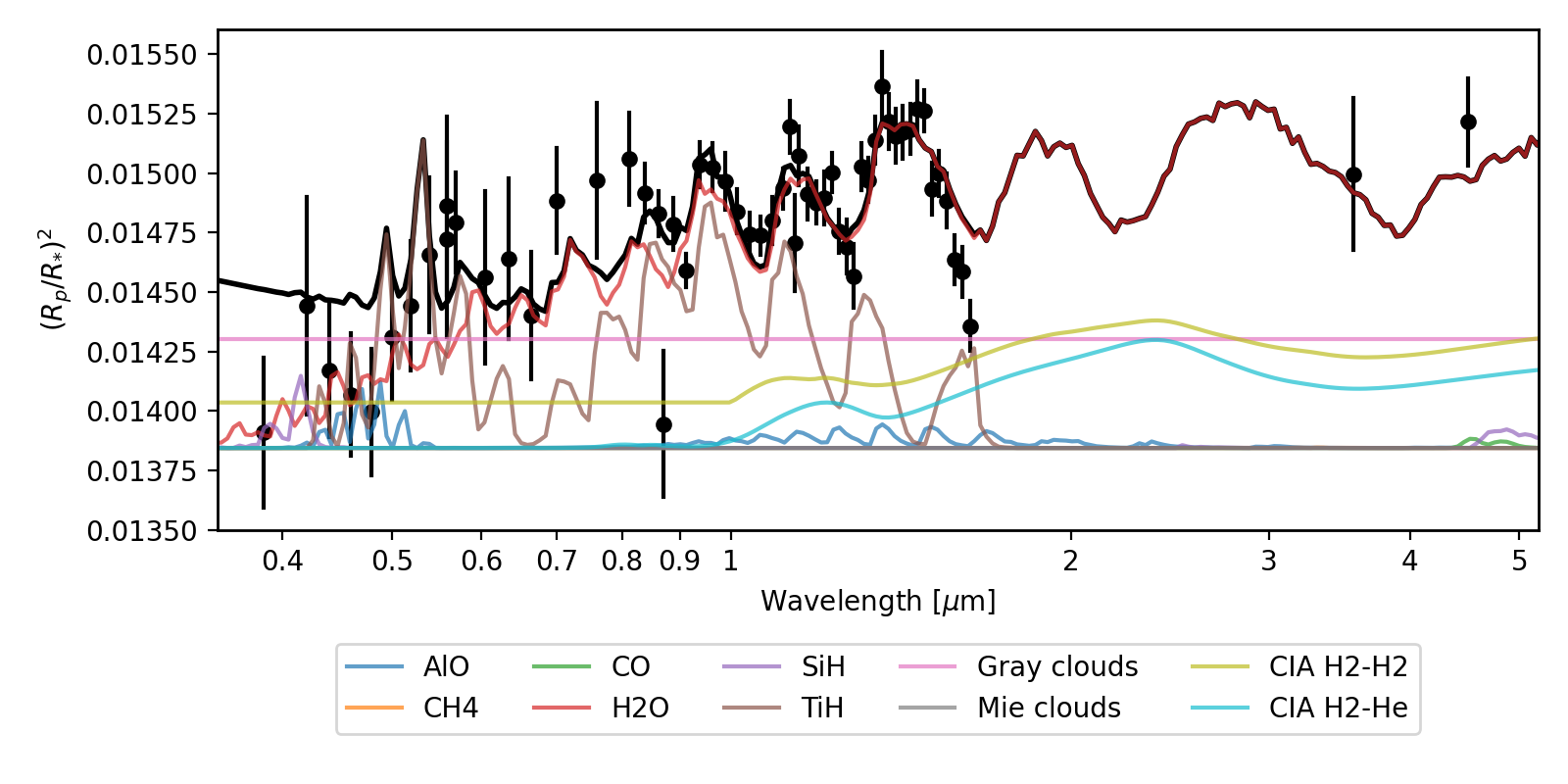}
    \caption{STIS visit 2\&3, WFC3 and Spitzer data (Case 2) fitted by the best-fit model (black line) that includes a homogeneous star together with the contributions from CIA, active trace gases, clouds and hazes.}
    \label{fig:scenario2_plus_contributions_no_stellar}
\end{figure*}

\end{document}